\documentclass[%
reprint,
superscriptaddress,
 amsmath,amssymb,
 aps,
 prb,
]{revtex4-2}

\usepackage{graphicx}% Include figure files
\usepackage{dcolumn}% Align table columns on decimal point
\usepackage{bm}% bold math
\usepackage[normalem]{ulem}
\usepackage{xcolor}
\newcommand{\pd}{\partial}
\renewcommand{\t}[1]{{\text{#1}}}

\begin{document}

\title{A tunable magneto-acoustic oscillator with low phase noise }% Force line breaks with \\

\author{A. Litvinenko} \thanks{Currently with Spintec, France. Correspondence to: LitvinenkoAN@gmail.com}
\affiliation{Laboratory of metamaterials,
Saratov State University, 410012, Saratov, Russia.}

\author{R. Khymyn}
\affiliation{Department of Physics, University of Gothenburg, 412 96,
Gothenburg, Sweden.}

\author{V. Tyberkevych}
\affiliation{Department of Physics, Oakland University,
48309, Rochester, Michigan, USA.}

\author{V. Tikhonov}
\affiliation{Laboratory of metamaterials,
Saratov State University, 410012, Saratov, Russia.}

\author{A. Slavin}
\affiliation{Department of Physics, Oakland University,
48309, Rochester, Michigan, USA.}

\author{S. Nikitov}
\affiliation{Laboratory of metamaterials,
Saratov State University, 410012, Saratov, Russia.}
\affiliation{Kotelnikov Institute of Radioengineering and Electronics of Russian Academy of Sciences, 125009, Moscow, Russia.}
\affiliation{Moscow Institute of Physics and Technology (National Research University), 141700, Dolgoprudny, Moscow Region, Russia.}


\date{\today}% It is always \today, today,

\begin{abstract}

A frequency-tunable low phase noise magneto-acoustic resonator is developed on the base of a parallel-plate straight-edge bilayer consisting of a yttrium-iron garnet (YIG) layer grown on a substrate of a gallium-gadolinium garnet(GGG). When a  YIG/GGG sample forms an ideal parallel plate, it supports a series of high-quality-factor acoustic modes standing along the plate thickness. Due to the magnetostriction of the YIG layer the ferromagnetic resonance (FMR) mode of the YIG layer can strongly interact with the acoustic thickness modes of the YIG/GGG structure, when the modes' frequencies match. A particular acoustic thickness mode used for the resonance excitations of the hybrid magneto-acoustic oscillations in a YIG/GGG bilayer is chosen by the YIG layer FMR frequency, which can be tuned by the variation of the external bias magnetic field. A composite magneto-acoustic oscillator, which includes an FMR-based resonance pre-selector, is developed to guarantee satisfaction of the Barkhausen criteria for a single-acoustic-mode oscillation regime.  The developed low phase noise composite magneto-acoustic oscillator can be tuned from 0.84 GHz to 1 GHz with an increment of about 4.8 MHz (frequency distance between the adjacent acoustic thickness modes in a YIG/GGG parallel plate), and demonstrates the phase noise of -116 dBc/Hz at the offset frequency of 10 KHz.

\begin{description}
%\pacs{85.70.Ec}% PACS, the Physics and Astronomy
\item[PACS numbers: 85.75.-d, 05.45.Xt, 75.40.Gb, 75.47.-m, 84.30.Qi]

%Use showkeys class option if keyword
\end{description}

\end{abstract}

\maketitle

%\tableofcontents

\section{\label{sec:level1}Introduction}
One of the most important tasks in the modern communication and radar technology is the development of reference oscillators with low phase noise, as the low level of phase noise translates into a high level of frequency stability necessary for the improved device performance. Also, in digital communication systems phase noise affects the system bit-error rate, and, therefore, the speed of data processing. In radar applications, lowering the phase noise leads to the increase of a radar range and sensitivity, as it allows to detect a signal reflected from the target with a lower power level.

In many common applications, reference or local \emph{tunable} oscillators are based on the yttrium-iron garnet (YIG) resonators, because the frequency of a ferromagnetic resonance (FMR) in YIG can be easily tuned over a decade by applied bias magnetic field. Also YIG resonators biased by powerful permanent magnets could have rather high resonance frequencies lying in the GHz frequency range, and demonstrate a relatively low linewidth, and, therefore, a relatively low level of the phase noise, especially at the reasonably large offset frequencies from the carrier. Another common method to reduce the oscillator phase noise is to use \emph{voltage controlled oscillators} (VCO) stabilized with a phase locked loop (PLL) \cite{Carter1984,Chenakin2011,Best2007}, but, although this  technique allows to significantly reduce the close-in phase noise, the far-out phase noise still remains determined by the intrinsic parameters of the used VCO.

The phase noise of an oscillator can be estimated using an empirical Leeson's equation\cite{Leeson1966}:
\begin{equation}\label{eq:1} %Lesson's equation
\mathcal{L}(\Delta\omega)=10log\bigg[\frac{FkT}{2P_{s}}\bigg(1+\Big(\frac{\omega_0}{2Q\Delta\omega}\Big)^2\bigg)\bigg(1+\frac{\omega_c}{\Delta\omega}\bigg)\bigg]
\end{equation}
where $\omega_0$ is the oscillator central (or "carrier") frequency, $\Delta\omega$ –- is the  "offset" frequency, $P_{s}$ – is the  signal power, $F$ – is the noise factor of the oscillator active element, $k$ – is the Boltzmann constant, $T$ – is the ambient absolute temperature, $Q$ – is the unloaded resonator quality factor, and $\omega_c$ – is the flicker corner frequency \cite{Leeson1966}. As it follows from the Leeson's equation (\ref{eq:1}), both the "close-in" and the "far-out" levels of the phase noise of an oscillator are, mainly, determined by the quality factor of an resonator used in the oscillator.

Thus, the enhancement of the resonator $Q$-factor is a key element in the development of new reference oscillators  for information and signal processing\cite{Vorobiev2010,Geerlings2012,Xu2009,Cai2015}. This goal, in principle, can be achieved by using resonators with low energy losses, such as dielectric\cite{kajfez1986dielectric}, optoelectronic\cite{volyanskiy2008applications}, acoustic\cite{Lakin1993}, and magnetic oscillators \cite{Bankowski2015} or the combinations of these oscillator types\cite{VitkoOptoElec2018}.

\begin{figure*}[hbt!]
\includegraphics[width=17.5cm]{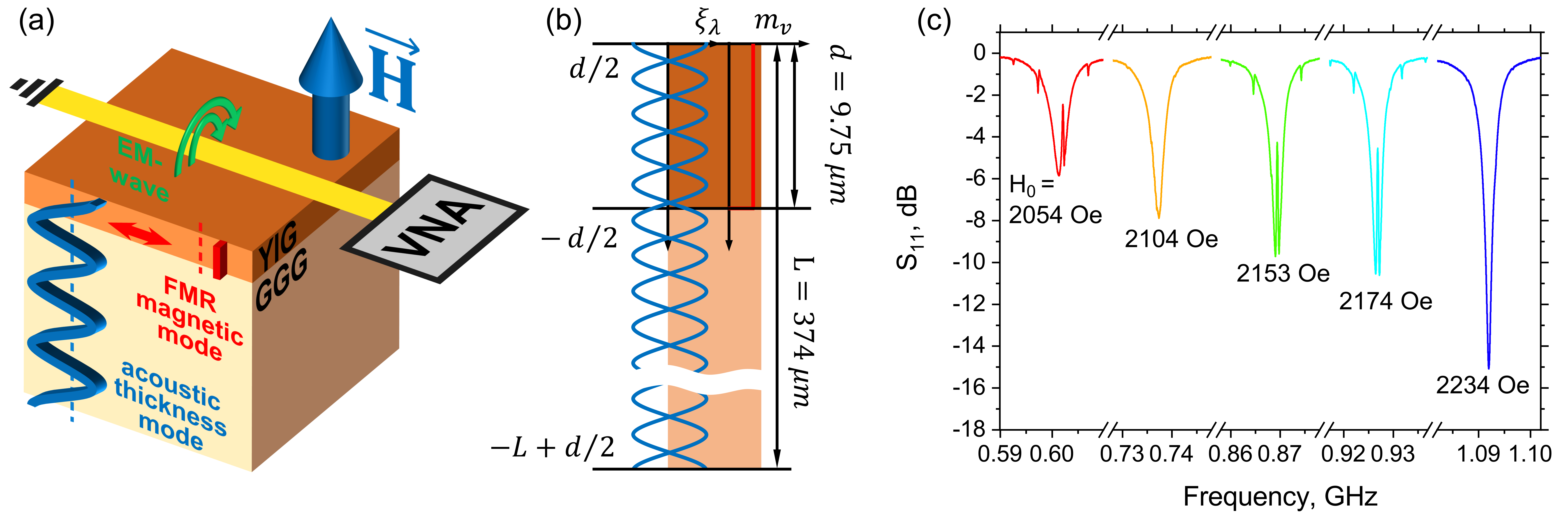}
\caption{a) Scheme of the a simple one-port reflection-based MAR , which was experimentally characterized using a vector network analyzer (VNA); b) Thickness distributions  of the magnetic FMR mode and standing acoustic modes in the  YIG/GGG bilayer sample; c) $S_{11}$-parameters of the one-port MAR at different values of the perpendicular-to-plane magnetic bias field.}
\label{figMAR} %~\cite{fig1}
\end{figure*}

The highest $Q$-factor, so far, is found in optoelectronic and dielectric resonators, but, unfortunately, these resonator types are, usually, rather bulky and have insufficient thermal stability of their resonance frequency. An alternative is to use the  solid-state acoustic resonators that can demonstrate $Q$-factors that are much higher than in magnetic YIG magnetic resonators,  while having sizes that are much smaller than the sizes of dielectric and optoelectronic resonators. Unfortunately, the purely acoustic resonators are not tunable.

A compromise solution would be to use hybrid magneto-acoustic resonators (MAR) that can support hybrid magneto-elastic oscillation modes that combine a high quality factor of the purely acoustic modes with the excellent tunability of the magnetic modes. It was shown in 1950s-60s that YIG has a considerable magnetostriction constant\cite{Comstock1965}, and that magneto-elastic waves of the GHz frequency range can be efficiently excited in magnetic layered films and hetero-structures \cite{Kittel1958, Spencer1958, Eshbach1963, Schlomann1964, Strauss1965, Auld1968, Rezende1969, Matthews1969}. In the 1980s the  technological progress in the liquid-phase epitaxy resulted in the development of high-quality (FMR linewidth below 0.5 Oe) YIG films grown on the (almost lattice-matched) mono-crystalline gadolinium gallium garnet (GGG) substrates. It was also demonstrated that magnetic oscillations excited in YIG through magnetostriction can effectively
excite standing acoustic thickness modes in the whole YIG-GGG garnet structure, because the sound velocities in YIG and GGG are almost equal \cite{Gulyaev1981, Kazakov1983, Zilberman1985, Gulyaev1988}. The interest to magneto-acoustic effects in garnet hetero-structures has been recently revived in a number of papers where YIG-GGG structures were used either in the transmission line configuration \cite{Weiler2012, Chowdhury2015, Khivintsev2018} or with ZnO  acoustical transducers which were used for a broad-band excitation of acoustic modes in these structures \cite{Polzikova2013, Pyataikin2017, Polzikova2019, Alekseev2020}.

Below, we show that a traditional parallel-plate straight-edge  YIG/GGG resonator can be successfully used as a tunable high-Q-factor magneto-acoustic resonance element of a local oscillator with a low phase noise. YIG/GGG films were previously used as hybrid magneto-acoustic resonators (MAR)\cite{Litvinenko2015, An2020}, where the YIG film served as an effective, narrow-band and frequency-tunable transducer which can selectively excite an acoustic thickness standing mode of the YIG/GGG structure, having a desirable frequency. In the configuration presented in Fig.~\ref{figMAR} the whole YIG/GGG structure acts as an effective high-overtone bulk acoustic resonator (HBAR)\cite{Boudot2016, Yu2009, Lakin1993}. Note, that HBARs among all the known acoustic resonators demonstrate the highest Q-factor (up to $10^{14}$), making the proposed YIG/GGG MAR design well-suited for the realization of low phase noise local oscillators. In this work, using the results of theoretical analysis of the magneto-acoustic interaction and experimental parameters of the YIG/GGG epitaxial parallel-plate structures, we design a novel tunable magneto-acoustic oscillator that has a level of phase noise, that is much lower than in conventional magnetic oscillators based only on the FMR mode of a YIG film.

\section{\label{sec:level2}Magneto-acoustic resonator with a high Q-factor}
The scheme of the YIG/GGG MAR is shown in the Fig.~\ref{figMAR}(a). It consists of a parallel-plate straight-edge rectangular resonator cut from a monocrystalline epitaxial YIG/GGG bilayer magnetized to saturation perpendicular to its plane by a bias magnetic field $\mathbf{H}_0$, and excited by a strip-line antenna connected to a vector network analyzer (VNA). The YIG film in the bilayer has the static magnetization $4\pi M_s$ = 1740 Gs , the FMR linewidth $\Delta H_0$ = 0.5 Oe and the thickness of 9.75 $\mu$m, and the in-plane sizes of 2x2 mm$^2$. The thickness of the GGG layer is 364 $\mu$m.

A signal  of a given frequency \emph{f} from the strip-line antenna excites the FMR mode in the YIG layer (uniform along the YIG film thickness) corresponding to a particular magnitude of the bias perpendicular bias magnetic field $H_0$. The FMR mode of the YIG resonator is coupled through the YIG magnetostriction to the standing thickness acoustic modes of the YIG/GGG hetero-structure, and  by simultaneous variation of the excitation frequency \emph{f} and the bias magnetic field $H_0$ it is possible to align any of the discrete thickness acoustic modes of the YIG/GGG hetero-structure with the YIG resonator FMR frequency given by the Kittel formula:
\begin{equation}
f=\gamma (H_0 - \mu_0 M_s),
\label{eq:FMRfreq}
\end{equation}
where $\gamma = 28.3$ GHz/T - is a gyromagnetic ratio of YIG.

In Fig.~\ref{figMAR}(c) $S_{11}$-parameter of the MAR is shown at different values of bias magnetic field $H_0$. A distinctive feature of $S_{11}$-parameter of the MAR is a dip with narrow inverse acoustic peaks. A broadband dip belongs to the FMR mode of the YIG film, while the narrow acoustic peaks which appear at bias field $H_0$ values of 2054, 2153 and 2174 Oe  correspond to the high overtone magnetoacoustic resonances in the YIG-GGG structure. Note, also that there are frequencies and corresponding bias field values at which acoustic peaks do not appear within the FMR dip. This indicates the absence of the magneto-acoustic coupling.

\begin{figure*}[hbt!]
\centering
\includegraphics[width=17.5cm]{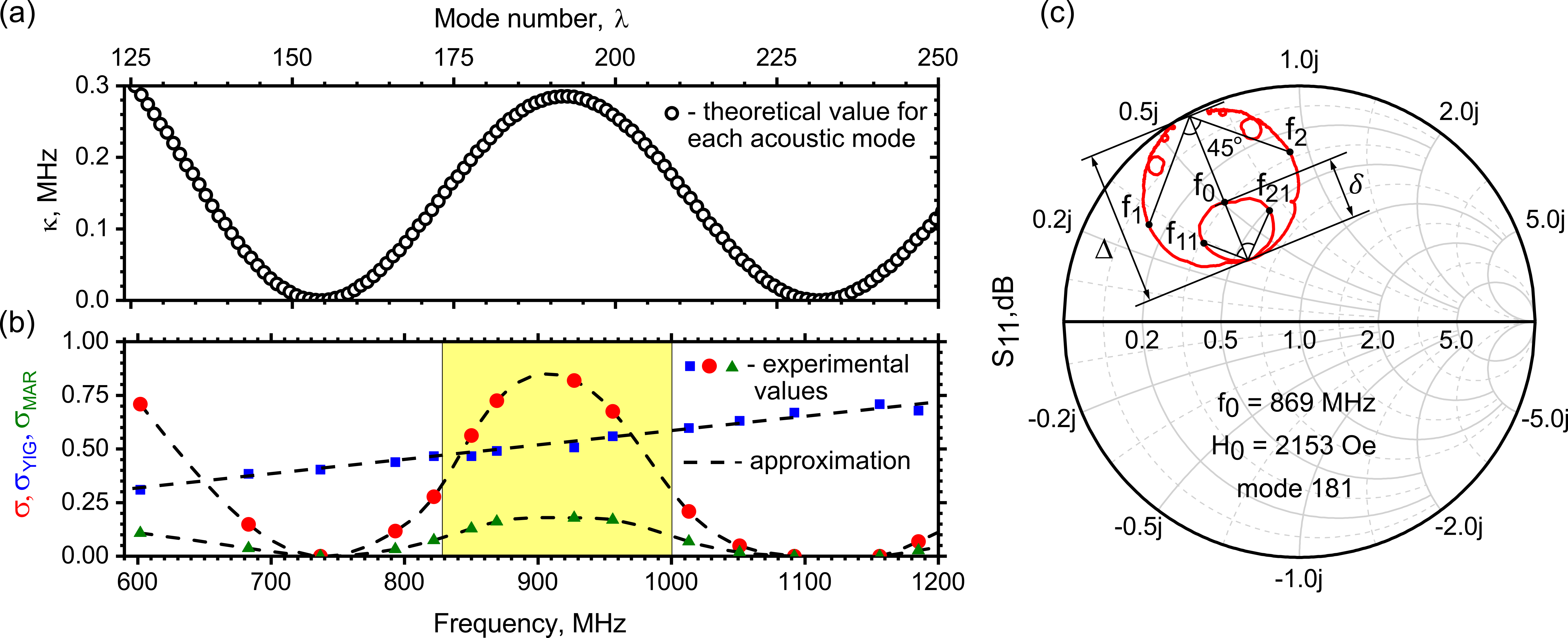}
\caption{Coupling coefficients in the MAR as functions of the excitation frequency and the excited acoustic mode number $\lambda$:
(a) - theoretically calculated coupling coefficient between the FMR mode of the YIG resonator and the thickness acoustic modes of the YIG/GGG structure; (b) - experimentally measured coupling  coefficients : $\sigma_{\t{YIG}}$ -- between the strip-line line and the FMR mode of the YIG resonator(blue squares); $\sigma$ -- between the FMR mode of the YIG resonator and the acoustic thickness modes of the YIG/GGG structure (red circles), $\sigma_\t{{MAR}}$ -- overall coupling between the strip-line and the acoustic thickness modes (green triangles). The region highlighted in yellow indicates the frequency band where the overall coupling coefficient is suitable for the operation of the  oscillator scheme. Note, that the FMR frequency of the YIG resonator is adjusted for a particular acoustic mode by changing the bias magnetic field $H_0$. In the frame (c) experimental $S_{11}$-parameter at the magnetic field $H_0=2153$ Oe and in the frequency band $869\pm15$ MHz  is plotted on the Smith charts. The main loop with a diameter $\Delta$ corresponds to the FMR mode,  while inner loops correspond to acoustic modes. The central inner loop with a diameter $\delta$ corresponds to the acoustic mode with which the FMR frequency is aligned. The values of  $\Delta, \delta$ are used to calculate the  experimental values of $\sigma_{\text{YIG}}$, $\sigma_{\text{MAR}}$, $\sigma$,  while the frequencies $f_0, f_1, f_2, f_{11}, f_{22}$ are used to calculate the Q-factors $Q_{\t{YIG}}, Q_{\t{MAR}}$, as described in the "Methods". At the frequencies where the coupling coefficient $\delta$ is close to zero, the inner acoustic loops disappear from the MAR $S_{11}$-parameter diagram.}
\label{figCouplingMAR}
\end{figure*}

To analyze the magneto-acoustic coupling in the proposed structure as a function of the frequency and the acoustic mode number we employ the theoretical description of the magneto-acouistic interaction in the sample (Fig.~1) developed in \cite{Verba2018, Lisenkov2019}. We start with the density of magneto-elastic energy in the form:
\begin{equation}
W=W_{\text{mag}}+W_{\text{el}}+W_{\text{mel}}.
\label{eq:mac_energy}
\end{equation}
 Here $W_{\text{mag}}$ is the density of magnetic energy , which includes contributions from the Zeeman, exchange and magneto-dipolar interactions; $W_{\text{el}}=\left[\rho\left(d \xi/dt\right)^2+c_{iklm}u_{ik}u_{lm}\right]/2$  is the elastic energy with the tensor of elastic constants $c_{iklm}$ and $u_{ik}=\left(d \xi_i/dx_k+d\xi_k/dx_i \right)/2$, $\xi$ is an acoustic displacement, and $\rho$ is the density of the material; $W_{\text{mel}}=b_{iklm}M_i M_k u_{lm}/M_s^2$ is the magneto-elastic interaction with  the  tensor  of  magnetostriction constants $b_{iklm}$\cite{chikazumi2009physics}.

Using the expression for the energy in Eq. (\ref{eq:mac_energy}), one can write two coupled equations for the elastic displacement $\xi$ and magnetization $\mathbf{M}$ as:
\begin{align}
\rho \frac{\pd^2 \xi_m}{\pd t^2} &=\frac{\pd}{\pd x_l} \frac{\pd W}{\pd u_{lm}} =\frac{b_{iklm}}{M_s^2} \frac{\pd M_i}{\pd x_l}M_k+c_{iklm}\frac{\pd u_{ik}}{\pd x_l} \label{initeqs1} \\
\frac{d\mathbf{M}}{dt} &= -\gamma\left[\mathbf{M} \times \left(\mathbf{H}^{\text{mag}}+\mathbf{H}^{\text{mel}}\right)\right],
\label{initeqs2}
\end{align}

where $\mathbf{H}^{\text{mag}}=\pd W_{\text{mag}}/\pd \mathbf{M}$ and $H_{i}^{\text{mel}}=2b_{iklm}M_{k}u_{lm}/M_s^2$.

Below we represent the magnetization vector $\mathbf{M}$ as a sum of its  static and precessional (dynamic) parts,  and the latter is expressed as the FMR mode of the thin YIG film:
\begin{equation}
\mathbf{M}=M_s\left[\boldsymbol{\mu}_0+\mathbf{m}
a(t) e^{-(i \omega_{0}+\Gamma) t}+ c.c.\right],
\end{equation}
where $\omega_{0}$ and $\Gamma$ are the angular frequency and damping parameter of the FMR mode.

In our case the static magnetization of YIG $\boldsymbol{\mu}_0$ is perpendicular to the film plane, while the dynamic magnetization $\mathbf{m} \perp \boldsymbol{\mu}_0$ describes the spatial (thickness) profile of the FMR mode in the absence of the magneto-elastic interaction $\mathbf{H}^{ac}=0$. Similarly, we represent the dynamic acoustic displacement, using the known profiles of the acoustic thickness eigenmodes of the YIG-GGG structure:

\begin{equation}
\boldsymbol{\xi}(z)=\sum_\lambda \tilde{\boldsymbol{\xi}}_\lambda(z) b_\lambda (t)
e^{-(\tilde\Gamma+i \tilde\omega_{\lambda}) t} +c.c.
\end{equation}
Taking into account the following orthogonality relations for magnetic and acoustic modes
\begin{align}
\frac{M_s}{\gamma} \int_{-d/2}^{d/2} \mathbf{m^*}(\boldsymbol{\mu}_0 \times
\mathbf{m}) dz &=-i A \\
2 \rho \omega_{\lambda} \int_{-L+d/2}^{d/2} \tilde{\boldsymbol{\xi}}_\lambda^{*} \tilde{\boldsymbol{\xi}}_{\lambda'} dz &= Q_{\lambda} \delta_{\lambda \lambda'},
\end{align}
one can rewrite Eqs. (\ref{initeqs1}-\ref{initeqs2}) as:
\begin{equation}
\begin{array}{c}
A \left[\dot{a}(t)+ i \omega a(t) + \Gamma_{0} a(t) \right]= i \kappa_{\lambda} b(t) \\
B_\lambda \left[\dot{b}_\lambda(t)+ i \tilde\omega_\lambda b_\lambda(t) + \tilde\Gamma_{\lambda} b_\lambda(t) \right]= i \kappa_{\lambda}^{*} a(t),
\end{array}
\end{equation}
with the coupling constant defined by the expression:
\begin{equation}
\kappa^2_{\lambda}=\frac{b^2\gamma}{2\omega_\lambda \rho M_s L d}
\left|\int \mu_{0} \mathbf{m}^{*} \frac{\pd \tilde{\boldsymbol{\xi}}_{\lambda}}{\pd z} dz\right|^2,
\end{equation}
where $b=b_{1111}-b_{1122}$.  If $\mathbf{m}$ and $\mathbf{\xi}$ describe plane waves (i.e. magnons and phonons), the coefficient $\kappa$ defines the bandgap at the point of avoided crossing( hybridization) of their spectra. In our case $\kappa^2$ defines the part of the oscillator "energy" involved in the magneto-elastic interaction. Please, note that the interaction coefficient $\kappa$  is expressed in the units of frequency, i.e. it is defined in relation to the central (carrier) frequency of the MAR.

We assume that the thickness profiles of both the FMR magnetic mode and the standing acoustic modes satisfy the "free" (or "unpinned") boundary conditions at the both parallel-plate surfaces and at the YIG-GGG interface:
\begin{equation}
\begin{array}{c}
\tilde{m}=1 \\
\tilde{\xi}_\lambda(z)=\cos\left[\left(z-d/2\right)\pi \lambda/L\right],
\end{array}
\end{equation}
For the spatially uniform static magnetization in the YIG layer with a sufficiently sharp transition at the  YIG-GGG  interface one can write
$\mu_0(z)=\Theta(d/2+z) \Theta(d/2-z)$ and
$m(z)=\tilde{m} \Theta(d/2+z) \Theta(d/2-z)$, where $\Theta(z)$ means Heaviside theta function.

Finally, for the coupling coefficient\cite{brataas2020spin} we obtain:
\begin{equation}
\kappa^2_\lambda=\frac{\gamma b^2}{2 \pi^2 d \lambda
	M_s \sqrt{c_{44} \rho}
}\left(1-\cos\frac{\pi \lambda d}{L}\right)^2, \label{kappaFMR}
\end{equation}
where $c_{44}$ is the elastic modulus of YIG\cite{borovik2012spin}.

The theoretically calculated coupling coefficient for the YIG/GGG sample used in our experiments is shown in Fig.2 (a) (black open circles). As it can be seen from the figure, the coupling coefficient $\kappa$ demonstrates an oscillating behavior: the overlap integral between the thickness profiles of the FMR mode and the acoustic modes has  local maxima when there are $n/2$ acoustic wavelength over the thickness of the YIG film,  and this integral vanishes to zero when there are $(n+1)/2$ acoustic wavelength over the thickness of the YIG layer. The oscillations in the magnitude of the coupling coefficient  reduce the operating range of frequencies for the MAR. As it follows from the theory,  in order to increase the period of oscillations of the coupling coefficient $\kappa$ and extend the operating frequency range of the MAR one has to reduce the thickness of the YIG film.

The magneto-elastic coupling can be described using a Darko Kajfez's method\cite{Kajfez1984} modified for hybrid MAR having two resonant subsystems (see description in the  "Methods" section). The experimentally measured oscillations in the magneto-elastic coupling have the same period as the ones calculated theoretically. This agreement between the theory and the experiment confirms that in the proposed structure of a MAR the  FMR mode, indeed, excites the acoustic shear modes of the YIG-GGG structure.

Due to the  relatively high value of YIG resonator Q-factor $Q_{\t{YIG}}\approx200-400$ the frequency bandwidth of the FMR in a laterally constrained  YIG resonator is narrower, than the frequency spacing between the acoustic thickness modes of the YIG-GGG structure $\Delta f_{a} = V_{a}/(2L) = 4.773\t{MHz}$, where $V_{a} = 3.57\times10^5 \t{cm/s}$ is the velocity of transverse acoustic waves in GGG (for comparison, the transverse acoustic wave velocity in YIG is $V_{a} = 3.85\times10^5 \t{cm/s}$). With this the YIG-film resonator can selectively excite a single acoustic shear mode of the YIG-GGG structure without using any narrow-band external filters. A YIG-GGG MAR can be tuned to excite effectively a single acoustic resonance mode having numbers from 173 to 208 in the [840MHz : 1.0GHz] frequency band with the step $\Delta f_{a}=4.8\t{MHz}$ by changing the magnitude of the bias magnetic field $H_0$ applied to the YIG film.

Since the YIG film works as a transducer between the electric signals in the strip-line antenna  and the acoustic thickness modes, the overall coupling coefficient between strip-line and acoustic modes has to be taken into account for the oscillator design. The overall coupling coefficient is shown in the Fig.~\ref{figCouplingMAR}(b). The detailed description on how to obtain experimental coupling coefficients of the hybrid YIG-GGG resonator is given in the "Methods".

\section{\label{sec:level3}Design of magneto-acoustic oscillator}
An important parameter for the design of an oscillator which employs high overtone resonators is the mode selectivity. In order to get stable oscillations without modulations and random spurs in the phase noise characteristic one has to make sure that when a particular mode is selected to be resonant, the damping of the adjacent modes is sufficiently strong. For the above described  MAR, the selectivity depends on the frequency separation between acoustic modes and on the linewidth of the FMR mode of the YIG layer. 

Using our experimental data, we found that the coupling coefficient of the MAR with the resonant acoustic mode having number 182 is 3.5 times larger than the corresponding coupling coefficients with the  adjacent acoustic  modes having numbers 181/183. The caulking advantage for the resonant acoustic mode can be further increased if an additional YIG-preselector is used. Moreover, with the increase of the central frequency of the MAR the loaded Q-factor of the magnetic (YIG) resonator grows linearly and, therefore, the effective linewidth of the YIG FMR mode decreases, thus increasing selectivity of the MAR.

\begin{figure}[ht]
\centering
\includegraphics[width=8cm]{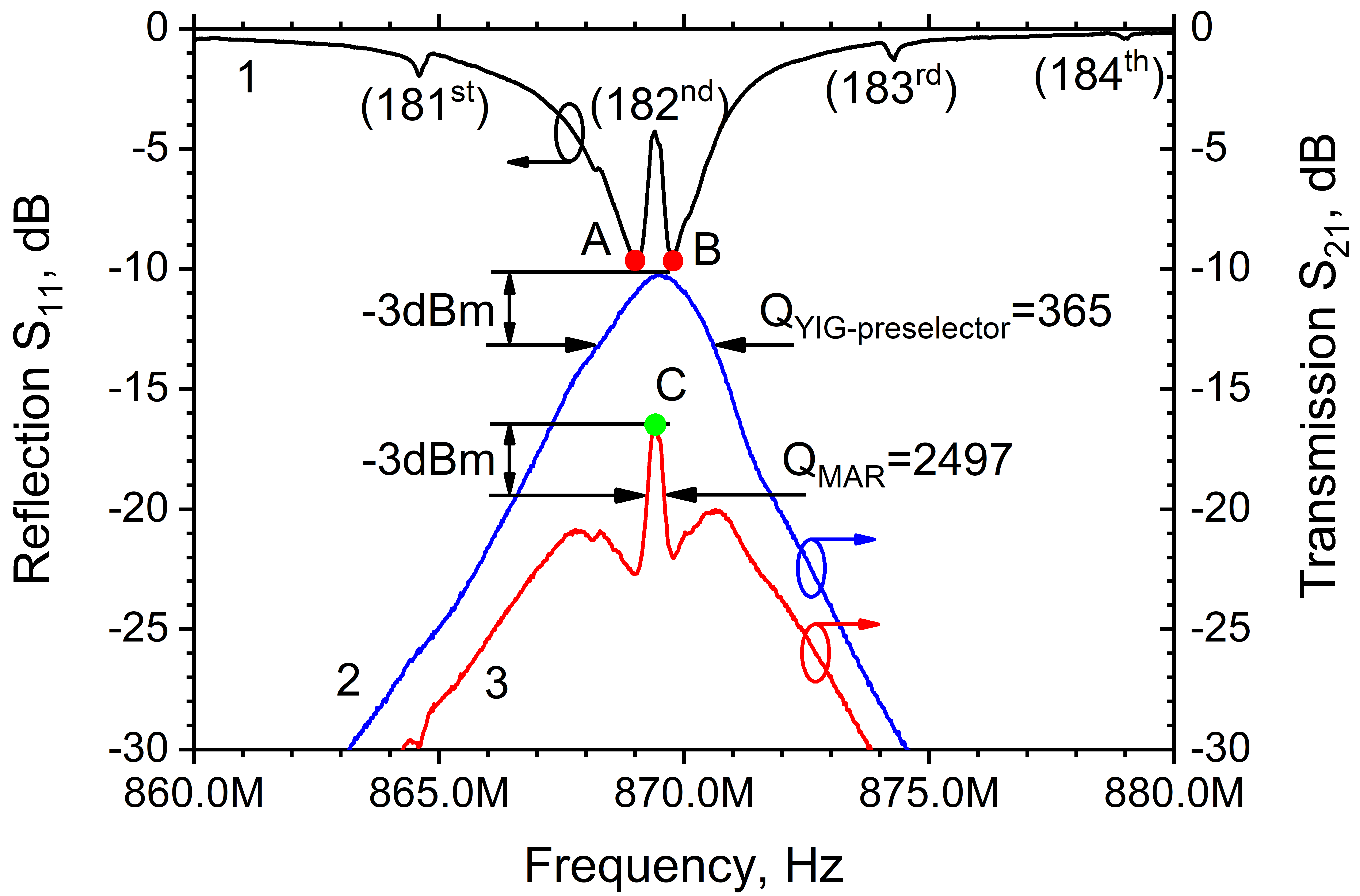}
\caption{S-parameters of the one-port (see Fig.1) and two-port composite (Fig.4) MARs: black line - $S_{11}$-parameter of the one-port MAR and $S_{21}$-parameter of a composite two-port MAR; blue line - $S_{21}$-parameter of the FMR-based preselector; red line - $S_{21}$-parameter of the composite two-port MAR. The mode numbers of a corresponding acoustic thickness modes in the YIG/GGG parallel plate are given in brackets. A, B and C are the points where the  Barkhausen criterion for stable auto-oscillations is satisfied}
\label{figSparameters}
\end{figure}

The basic principles of the design of an efficient  magneto-acoustic oscillator can be understood from the analysis of the experimental S-parameter data presented in Fig.~\ref{figSparameters}. 
First of all, let us look at the experimentally measured overall coupling coefficient of the one-port simple MAR presented in Fig.1 (green triangles in Fig.~\ref{figCouplingMAR}). In the experience of practical oscillator design, the minimal coupling coefficient between the oscillator core and the resonator should be above 0.1 for high stability and low phase noise. Therefore, according to this empirical condition the oscillator based on the MAR can operate in the range of frequencies between 0.84 and 1GHz (see the region highlighted in yellow in Fig.\ref{figCouplingMAR}). For further design steps we chose the resonance acoustic mode within a yellow region having number 182, and the Q-factor of $Q=2497$. As it was discussed earlier, the resonance characteristic of a one-port reflection-based MAR has an unusual form of a dip, caused by the FMR in the YIG layer with an inverted central peak in the middle attributed to the resonance acoustic mode with the number 182 of the whole YIG-GGG structure. The analysis, however, shows that if we use a conventional one-port reflection-based oscillator design, the Barkhausen stability criterion for the auto-oscillations is only satisfied at the points A and B (see Fig.~\ref{figSparameters} which are situated outside of the central peak of the acoustic resonance for the mode 182. Therefore, the use of such a design for a magneto-acoustical oscillator (MAO) will not substantially decrease the MAO phase noise figure, since the phase noise will, mostly, be determined  by a relatively low Q-factor of the FMR mode.  Moreover, in the systems where several competing resonance modes (corresponding to points A and B on the black curve) can be excited simultaneously it is possible to have mode bistability and chaotic dynamics\cite{litvinenko2018chaotic}.

In order to take the full advantage of the high Q-factor of a \emph{single acoustic resonance mode}, a special scheme based on the one-port MAR is designed to satisfy with the Barkhausen stability criterion near the frequency of the acoustic resonance mode. It is done in two steps. First, a circulator is added serially to the one-port MAR forming a two-port circuit which has an $S_{21}$-parameter absolutely identical to a $S_{11}$-parameter of the one-port MAR. For a ring oscillator scheme based on such a two-port circuit (see a part of the scheme between point 1 and 3) given adjusted open loop phase and amplification the Barkhausen stability criterion would be satisfied at the same time at the narrow peak corresponding to the frequency of the acoustic resonance and at frequencies far from the FMR dip. Therefore, in a second step we introduce an additional purely magnetic two-port YIG-resonator patterned on the same GGG substrate as a pre-selector-bandpass-filter (having a \emph{scratched bottom GGG surface} to prevent the formation of the standing acoustic thickness modes) which suppresses the signal at frequencies outside of the FMR resonance. As a result of such a design, the transmission ($S_{21}$) characteristic of the composite two-port MAR (red line in Fig.3) is approximately a product of the transmission characteristic of the two-port circuit based on the MAR and a circulator (black curve in Fig.3), and the transmission characteristic of the FMR-based pre-selector (blue curve in Fig.3). As a result the transmission characteristic of the composite two-port MAR (Fig.4) has a usual form of a resonance characteristic with a central maximum. We note, that a MAR transmission characteristic with a central maximum, similar to the one shown by the  red curve in Fig.3, can be obtained by simpler means in a three-layer YIG-GGG-YIG structure which was used in \cite{An2020}, without use of an additional pre-selector. However, in that case the thickness of the GGG layer can not be adjusted by polishing, and the frequency spacing of the acoustic thickness modes mode can not be adjusted after the growth of the YIG layers by liquid epitaxy. Another limitation is that the liquid epitaxy process requires the GGG substrate thickness to be at least 300 $\mu$m, while with polishing of one sided YIG-GGG structure this thickness can be reduced down to 50-100 $\mu$m.

\begin{figure}[ht]
\centering
\includegraphics[width=8cm]{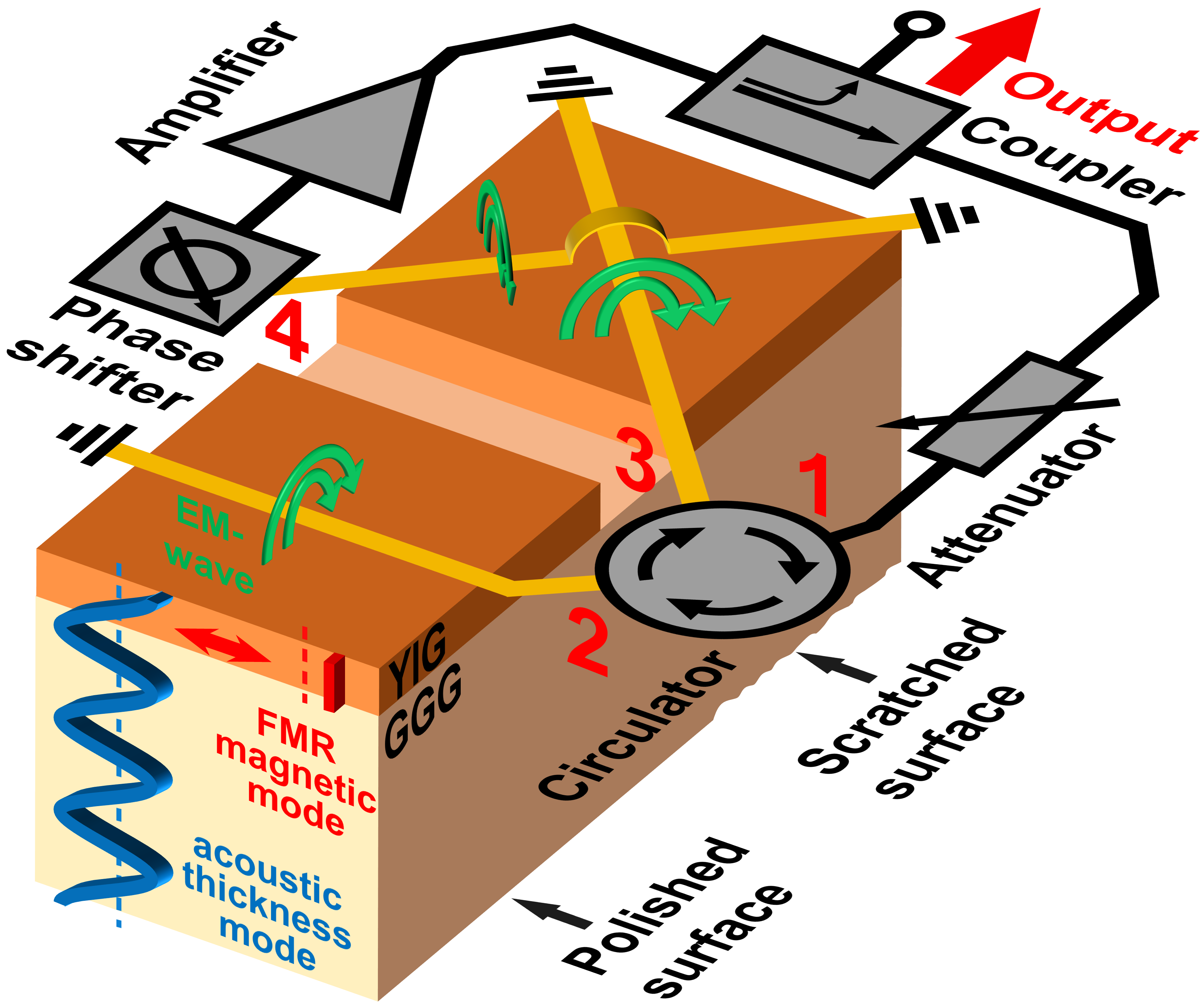}
\caption{Scheme of a composite two-port MAR consisting of a YIG/GGG MAR (left part of a YIG layer) connected to a YIG FMR-based pre-selector-filter (right part of the YIG layer). The left part forms a two-port MAR , similar to the one shown in Fig.1, connected to the purely magnetic YIG-FMR-based pre-selector-filter formed by the right part of the YIG layer, where the acoustic modes are eliminated by  scratching the bottom surface of the GGG substrate.}
\label{figMAO}
\end{figure}

Finally, to complete the scheme of MAO a low phase noise amplifier ABA-54563 is added (see the circuit Fig.4) to compensate losses in the oscillator feedback loop. A variable delay line is used to obtain the correct phase shift in the loop to satisfy the phase condition of the Barkhausen stability criterion for oscillations. A variable attenuator is introduced to limit the signal power at the input of the MAR to suppress the  nonlinearities and avoid the increase of the phase noise. Finally, a coupler is used to extract the output signal. Given the adjusted amplification, phase shift in the feedback loop, and the bias magnetic field applied to the YIG film, the auto-oscillation conditions for the composite two-port MAR are satisfied only at the frequency of the acoustic thickness mode having the number 182 and the effective Q-factor of 2497 (see point "C" in Fig.~\ref{figSparameters}). Moreover, the adjacent acoustic thickness mode with the number 181  is suppressed  by 13~dB as compared  to the resonant mode with the number 182.  Thus,  a single-mode auto-oscillation regime based on a high-Q-factor acoustic thickness mode in a composite two-port MAR is realised.  

\section{\label{sec:level4}Results on low phase noise}
\begin{figure}[h]
\centering
\includegraphics[width=8cm]{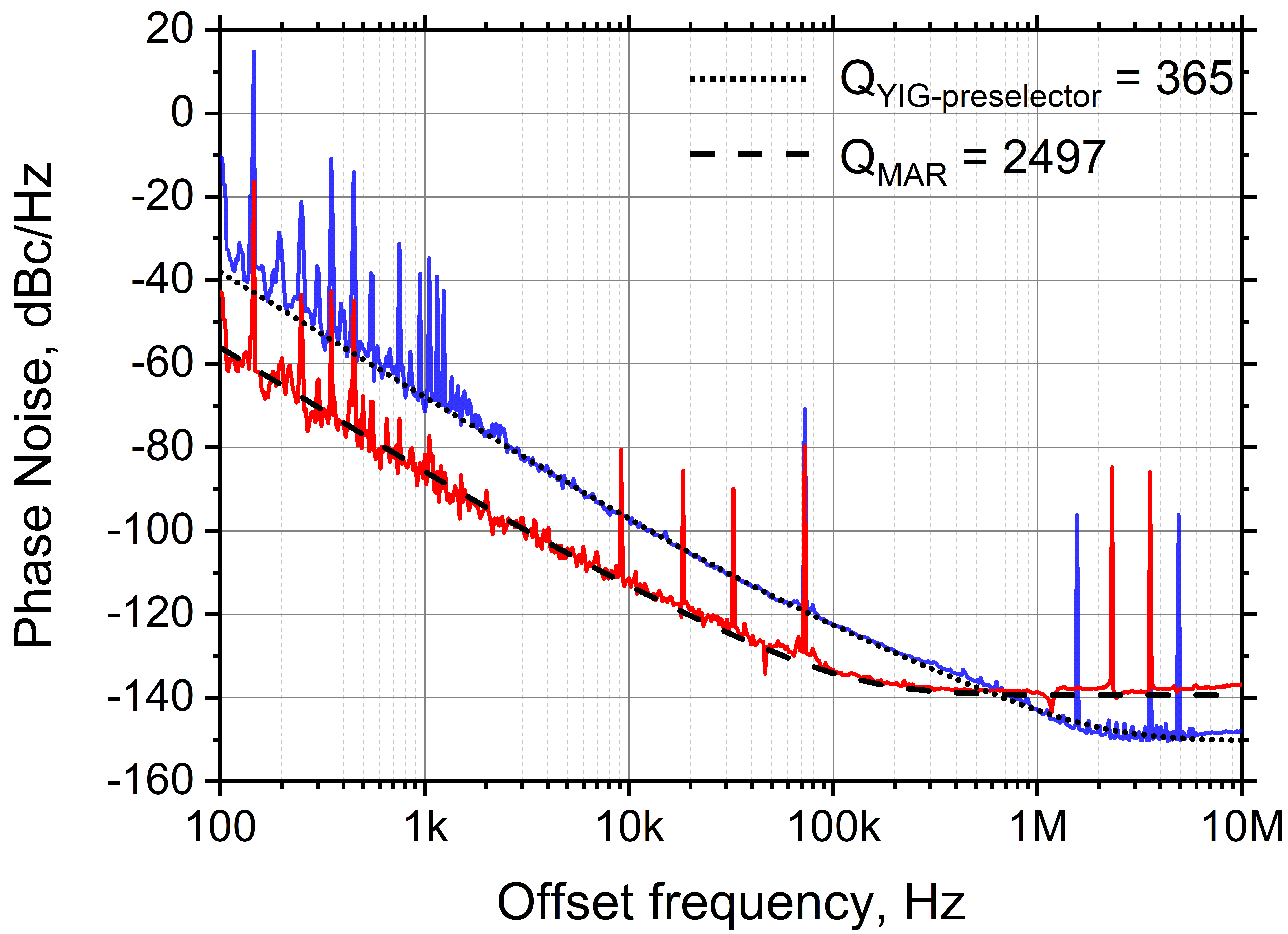}
\caption{Experimentally measured phase noise figures for auto-oscillators  based on a purely magnetic YIG FMR resonator (blue line)  and on a composite two-port magneto-acoustic YIG/GGG resonator (red line). Dotted black line shows the estimation of the phase noise obtained from the Leeson's formula (1) for a FMR-based magnetic pre-selector resonator ($ Q= 365$) , while the dashed black line shows a similar estimation of a phase noise  for the two-port MAR (Fig.4) having the Q-factor $Q= 2497$.}
\label{figPhaseNoise}
\end{figure}

The experimentally measured phase noise figure of an auto-oscillator based  the composite two-port MAR (see Fig.4) having ($ Q= 2497$) is presented in Fig.~\ref{figPhaseNoise} by a red curve. In the same figure, for comparison, we show by the blue curve the experimentally measured phase noise figure for an auto-oscillator based on a purely magnetic YIG FMR pre-selector oscillator ($Q=365$) . The theoretical estimations of the phase noise figures in these two auto-oscillators obtained from Eq.(1) are shown in Fig.~\ref{figPhaseNoise} by the black dashed line and the black dotted line, respectively. 
 
The phase noise of the MAO based on the composite two-port MAR is -87dBc/Hz at the  1kHz offset and -116dBc/Hz at the  10kHz offset,  which is at least 20 dB better than the phase noise figure of a conventional auto-oscillator based on a YIG FMR resonator. Note, at the same time, that at large off-set frequencies the phase noise of an auto-oscillator based on a composite two-port MAR degrades, and becomes higher than in the case  of a conventional YIG-FMR oscillator. This increase in the MAR phase noise is caused, mainly, by the presence of a variable attenuator in the MAR scheme (Fig.~\ref{figMAO}) which is introduced to avoid saturation of the one-port MAR in nonlinear mode. Note, also, that the nonlinearity threshold of the MAR is lower, than of the YIG-FMR two-port resonator. We would like to mention that an amplifier with a lower saturation power level can be used instead  of the variable attenuator in the MAO design scheme to reduce the maximum level of power at the MAR input, and, consequently, the noise factor of the feedback loop. This will improve the phase noise figure in the whole range of offset frequencies.

\section{\label{sec:level5}Conclusion}
We have shown that the FMR mode excited in a YIG layer of a parallel-plate YIG/GGG hetero-structure can be effectively coupled to the high-Q-factor standing thickness acoustic modes of the YIG/GGG bilayer. The frequency dependence of this magneto-elastic coupling was studied both theoretically and experimentally, and the optimum conditions for this coupling corresponding to the stable single-mode auto-oscillations were found.  A composite two-port MAR based on the YIG/GGG hetero-structure was developed and practically realized, and the phase noise figure in the auto-oscillator based on the developed MAR was substantially improved, in comparison with the auto-oscillator based on the conventional YIG FMR two-port resonator. We would like to stress, that the advantage of using shear acoustic modes in the developed MAR over the longitudinal acoustic waves in conventional HBARs is caused by the fact that shear acoustic modes are insensitive to the amorphous load. This property provides a significant simplification of technological requirements for the manufacturing of the proposed composite MARs.

 In summary, we have designed a  composite magneto-acoustic auto-oscillator with a low phase noise based on a parallel-plate YIG/GGG bilayer. The relatively narrow FMR linewidth of the YIG layer provides the possibility for selective resonance excitation of a single acoustic thickness mode of the YIG/GGG structure and, therefore, makes possible a significant improvement of the Q-factor of the resulting MAR. The phase noise figure of the anto-oscillator based on the developed composite MAR  is -87dBc/Hz at 1kHz offset and -116dBc/Hz at 10kHz offset which is at least 20dB better than the performance of the conventional auto-oscillating scheme  based on a conventinal the YIG FMR two-port resonator. The designed MAO can be used in the frequency-agile data transmission where abrupt frequency-hoping is employed. As an outlook, we note that the microwave circulator used in our scheme of the composite two-port MAR can be replaced with a matching circuit or an active component to make the developed MAR  CMOS-compatible.

\begin{acknowledgments}
This work was supported by the Grant of the Government of the Russian Federation for supporting scientific research projects supervised by leading scientists at Russian institutions of higher education (Contract No. 11.G34.31.0030) and supported in part by the U.S. National Science Foundation (Grants No. EFMA-1641989), by the U.S. Air Force Office of Scientific Research under the MURI grant No. FA9550-19-1-0307, and by the Oakland University Foundation. The work of SAN was supported by Agreement with Ministry of Science and Higher Education of the Russian Federation $\#13.1902.21.0010$.

We would like to thank Dr. Olivier Klein for the help with precise measurement of the YIG and GGG thickness using interferometry technique.
\end{acknowledgments}

\appendix
\section{Device fabrication}

The scheme of the one-port reflection-type MAR is shown in the Fig.~1. The MAR is manufactured using a two layer parallel-plate YIG/GGG structure with the YIG thickness of $d=9.75\mu \t{m}$, GGG thickness of $L=364\mu \t{m}$, and YIG saturation magnetization of $M_{0} = 1740/(4\pi)Oe$.  The chemical-mechanical polishing technique was used to create the highly parallel surfaces of YIG/GGG structure with the wedge angle less than $2''$, which ensures the formation of high overtone acoustic thickness resonances in the YIG/GGG structure. The fabricated sample had the lateral dimensions of 1x2.5mm.

\section{Device characterization}
The developed MAR and the YIG pre-selector FMR resonator were characterized using Keysight (Agilent E8361A) vector network analyzer. The phase noise of the developed composite MAO was measured using a signal source analyzer (Keysight E5052B).

The experimental method used for the characterization of the magneto-acoustic coupling in the MAR was a modified Darko Kajfez's method. This method is based on the analysis of the resonance loops in an $S_{11}$-parameter graph plotted on a Smith chart. This method gives dimensionless values of the loaded Q-factors, and coupling coefficients that define interaction of electrical sub-systems in the filter theory. To obtain the $S_{11}$-parameter of the MAR we used an electrical setup shown in Fig.1a. The strip-line antenna used in our experiments had the width of 0.5 mm, which is comparable to the sample size, and ensures an efficient excitation of the FMR mode. In order to obtain the loaded Q-factors and coupling coefficients of the composite MAR some additional geometrical constructions were used on the Smith chart. First, to get the parameters of a magnetic resonator, three lines have to be drawn from the node of the big loop, as shown in Fig. 1c: one line goes through the center of the loop, while two other lines go at the angle of 45 degrees to each side of the center line. This allows us to extract from the S-parameter measurement the values of the corresponding frequencies $f_{1}$, $f_{2}$ and $f_{0}$ as the intersection points between the auxiliary lines and the S-parameter curve. The interval $f_{2}-f_{1}$ defines the FMR linewidth,  and is used to calculate the loaded Q-factor of the YIG magnetic resonator:
\begin{equation}
Q_{\t{YIG}}=\frac{f_{0}}{f_{2}-f_{1}},
\label{QfactorYIG}
\end{equation}
Similar geometric constructions were made with the largest inner loop, in order to obtain the frequencies $f_{11}$, $f_{21}$, which define the width of the resonant acoustic mode of the YIG/GGG structure, and its loaded Q-factor:
\begin{equation}
Q_{\t{MAR}}=\frac{f_{0}}{f_{21}-f_{11}},
\label{QfactorMAR}
\end{equation}
In order to calculate the coupling coefficients between the subsystems in the MAR one should measure the diameter of the main resonance loop and the largest inner loop. The coupling between the strip-line and the YIG FMR magnetic resonator is defined by the expression:
\begin{equation}
\sigma_{\t{YIG}}=\frac{\Delta}{2R-\Delta},
\label{SigmaYIG}
\end{equation}
where 2R=2 is a radius of the Smith Chart. The overall coupling between the strip-line and the acoustic subsystem can be obtained using a similar formula:
\begin{equation}
\sigma_{\t{MAR}}=\frac{\delta}{2R-\delta},
\label{SigmaMEC}
\end{equation}
In order to compare the theoretically calculated $\kappa$ with the experimental data we introduced a coupling coefficient $\sigma$,  which provides a measure of coupling between the magnetic and acoustic sub-systems:
\begin{equation}
\sigma=\frac{\delta}{\Delta-\delta},
\label{SigmaMAR}
\end{equation}
Detailed explanation and derivation of the above presented expressions can be found in \cite{Kajfez1984}.

\bibliography{references.bib}

\begin{thebibliography}{44}%
\makeatletter
\providecommand \@ifxundefined [1]{%
 \@ifx{#1\undefined}
}%
\providecommand \@ifnum [1]{%
 \ifnum #1\expandafter \@firstoftwo
 \else \expandafter \@secondoftwo
 \fi
}%
\providecommand \@ifx [1]{%
 \ifx #1\expandafter \@firstoftwo
 \else \expandafter \@secondoftwo
 \fi
}%
\providecommand \natexlab [1]{#1}%
\providecommand \enquote  [1]{``#1''}%
\providecommand \bibnamefont  [1]{#1}%
\providecommand \bibfnamefont [1]{#1}%
\providecommand \citenamefont [1]{#1}%
\providecommand \href@noop [0]{\@secondoftwo}%
\providecommand \href [0]{\begingroup \@sanitize@url \@href}%
\providecommand \@href[1]{\@@startlink{#1}\@@href}%
\providecommand \@@href[1]{\endgroup#1\@@endlink}%
\providecommand \@sanitize@url [0]{\catcode `\\12\catcode `\$12\catcode
  `\&12\catcode `\#12\catcode `\^12\catcode `\_12\catcode `\%12\relax}%
\providecommand \@@startlink[1]{}%
\providecommand \@@endlink[0]{}%
\providecommand \url  [0]{\begingroup\@sanitize@url \@url }%
\providecommand \@url [1]{\endgroup\@href {#1}{\urlprefix }}%
\providecommand \urlprefix  [0]{URL }%
\providecommand \Eprint [0]{\href }%
\providecommand \doibase [0]{https://doi.org/}%
\providecommand \selectlanguage [0]{\@gobble}%
\providecommand \bibinfo  [0]{\@secondoftwo}%
\providecommand \bibfield  [0]{\@secondoftwo}%
\providecommand \translation [1]{[#1]}%
\providecommand \BibitemOpen [0]{}%
\providecommand \bibitemStop [0]{}%
\providecommand \bibitemNoStop [0]{.\EOS\space}%
\providecommand \EOS [0]{\spacefactor3000\relax}%
\providecommand \BibitemShut  [1]{\csname bibitem#1\endcsname}%
\let\auto@bib@innerbib\@empty
%</preamble>
\bibitem [{\citenamefont {{Carter}}\ \emph {et~al.}(1984)\citenamefont
  {{Carter}}, \citenamefont {{Owens}},\ and\ \citenamefont
  {{De}}}]{Carter1984}%
  \BibitemOpen
  \bibfield  {author} {\bibinfo {author} {\bibfnamefont {R.~L.}\ \bibnamefont
  {{Carter}}}, \bibinfo {author} {\bibfnamefont {J.~M.}\ \bibnamefont
  {{Owens}}},\ and\ \bibinfo {author} {\bibfnamefont {D.~K.}\ \bibnamefont
  {{De}}},\ }\bibfield  {title} {\bibinfo {title} {Yig oscillators: Is a planar
  geometry better? (short papers)},\ }\href
  {https://doi.org/10.1109/TMTT.1984.1132912} {\bibfield  {journal} {\bibinfo
  {journal} {IEEE Transactions on Microwave Theory and Techniques}\ }\textbf
  {\bibinfo {volume} {32}},\ \bibinfo {pages} {1671} (\bibinfo {year}
  {1984})}\BibitemShut {NoStop}%
\bibitem [{\citenamefont {Chenakin}(2011)}]{Chenakin2011}%
  \BibitemOpen
  \bibfield  {author} {\bibinfo {author} {\bibfnamefont {A.}~\bibnamefont
  {Chenakin}},\ }\href@noop {} {\bibfield  {journal} {\bibinfo  {journal}
  {Microwaves and Rf}\ }\textbf {\bibinfo {volume} {50}},\ \bibinfo {pages}
  {72} (\bibinfo {year} {2011})}\BibitemShut {NoStop}%
\bibitem [{\citenamefont {Best}(2007)}]{Best2007}%
  \BibitemOpen
  \bibfield  {author} {\bibinfo {author} {\bibfnamefont {R.~E.}\ \bibnamefont
  {Best}},\ }\href@noop {} {\emph {\bibinfo {title} {Phase-Locked Loops}}}\
  (\bibinfo  {publisher} {New York:McGraw Hill},\ \bibinfo {year}
  {2007})\BibitemShut {NoStop}%
\bibitem [{\citenamefont {{Leeson}}(1966)}]{Leeson1966}%
  \BibitemOpen
  \bibfield  {author} {\bibinfo {author} {\bibfnamefont {D.~B.}\ \bibnamefont
  {{Leeson}}},\ }\bibfield  {title} {\bibinfo {title} {A simple model of
  feedback oscillator noise spectrum},\ }\href
  {https://doi.org/10.1109/PROC.1966.4682} {\bibfield  {journal} {\bibinfo
  {journal} {Proceedings of the IEEE}\ }\textbf {\bibinfo {volume} {54}},\
  \bibinfo {pages} {329} (\bibinfo {year} {1966})}\BibitemShut {NoStop}%
\bibitem [{\citenamefont {Vorobiev}\ and\ \citenamefont
  {Gevorgian}(2010)}]{Vorobiev2010}%
  \BibitemOpen
  \bibfield  {author} {\bibinfo {author} {\bibfnamefont {A.}~\bibnamefont
  {Vorobiev}}\ and\ \bibinfo {author} {\bibfnamefont {S.}~\bibnamefont
  {Gevorgian}},\ }\bibfield  {title} {\bibinfo {title} {Tunable thin film bulk
  acoustic wave resonators with improved q-factor},\ }\href
  {https://doi.org/10.1063/1.3441413} {\bibfield  {journal} {\bibinfo
  {journal} {Applied Physics Letters}\ }\textbf {\bibinfo {volume} {96}},\
  \bibinfo {pages} {212904} (\bibinfo {year} {2010})}\BibitemShut {NoStop}%
\bibitem [{\citenamefont {Geerlings}\ \emph {et~al.}(2012)\citenamefont
  {Geerlings}, \citenamefont {Shankar}, \citenamefont {Edwards}, \citenamefont
  {Frunzio}, \citenamefont {Schoelkopf},\ and\ \citenamefont
  {Devoret}}]{Geerlings2012}%
  \BibitemOpen
  \bibfield  {author} {\bibinfo {author} {\bibfnamefont {K.}~\bibnamefont
  {Geerlings}}, \bibinfo {author} {\bibfnamefont {S.}~\bibnamefont {Shankar}},
  \bibinfo {author} {\bibfnamefont {E.}~\bibnamefont {Edwards}}, \bibinfo
  {author} {\bibfnamefont {L.}~\bibnamefont {Frunzio}}, \bibinfo {author}
  {\bibfnamefont {R.~J.}\ \bibnamefont {Schoelkopf}},\ and\ \bibinfo {author}
  {\bibfnamefont {M.~H.}\ \bibnamefont {Devoret}},\ }\bibfield  {title}
  {\bibinfo {title} {Improving the quality factor of microwave compact
  resonators by optimizing their geometrical parameters},\ }\href
  {https://doi.org/10.1063/1.4710520} {\bibfield  {journal} {\bibinfo
  {journal} {Applied Physics Letters}\ }\textbf {\bibinfo {volume} {100}},\
  \bibinfo {pages} {192601} (\bibinfo {year} {2012})}\BibitemShut {NoStop}%
\bibitem [{\citenamefont {{Xu}}\ \emph {et~al.}(2009)\citenamefont {{Xu}},
  \citenamefont {{Zhang}}, \citenamefont {{Yu}}, \citenamefont
  {{Abbaspour-Tamijani}},\ and\ \citenamefont {{Chae}}}]{Xu2009}%
  \BibitemOpen
  \bibfield  {author} {\bibinfo {author} {\bibfnamefont {W.}~\bibnamefont
  {{Xu}}}, \bibinfo {author} {\bibfnamefont {X.}~\bibnamefont {{Zhang}}},
  \bibinfo {author} {\bibfnamefont {H.}~\bibnamefont {{Yu}}}, \bibinfo {author}
  {\bibfnamefont {A.}~\bibnamefont {{Abbaspour-Tamijani}}},\ and\ \bibinfo
  {author} {\bibfnamefont {J.}~\bibnamefont {{Chae}}},\ }\bibfield  {title}
  {\bibinfo {title} {In-liquid quality factor improvement for film bulk
  acoustic resonators by integration of microfluidic channels},\ }\href
  {https://doi.org/10.1109/LED.2009.2019973} {\bibfield  {journal} {\bibinfo
  {journal} {IEEE Electron Device Letters}\ }\textbf {\bibinfo {volume} {30}},\
  \bibinfo {pages} {647} (\bibinfo {year} {2009})}\BibitemShut {NoStop}%
\bibitem [{\citenamefont {Cai}\ \emph {et~al.}(2015)\citenamefont {Cai},
  \citenamefont {Lu}, \citenamefont {Chen}, \citenamefont {Lee}, \citenamefont
  {Lin},\ and\ \citenamefont {Yen}}]{Cai2015}%
  \BibitemOpen
  \bibfield  {author} {\bibinfo {author} {\bibfnamefont {D.-P.}\ \bibnamefont
  {Cai}}, \bibinfo {author} {\bibfnamefont {J.-H.}\ \bibnamefont {Lu}},
  \bibinfo {author} {\bibfnamefont {C.-C.}\ \bibnamefont {Chen}}, \bibinfo
  {author} {\bibfnamefont {C.-C.}\ \bibnamefont {Lee}}, \bibinfo {author}
  {\bibfnamefont {C.-E.}\ \bibnamefont {Lin}},\ and\ \bibinfo {author}
  {\bibfnamefont {T.-J.}\ \bibnamefont {Yen}},\ }\bibfield  {title} {\bibinfo
  {title} {High q-factor microring resonator wrapped by the curved waveguide},\
  }\href {https://doi.org/10.1038/srep10078} {\bibfield  {journal} {\bibinfo
  {journal} {Scientific reports}\ }\textbf {\bibinfo {volume} {5}},\ \bibinfo
  {pages} {10078} (\bibinfo {year} {2015})}\BibitemShut {NoStop}%
\bibitem [{\citenamefont {Kajfez}\ and\ \citenamefont
  {Guillon}(1986)}]{kajfez1986dielectric}%
  \BibitemOpen
  \bibfield  {author} {\bibinfo {author} {\bibfnamefont {D.}~\bibnamefont
  {Kajfez}}\ and\ \bibinfo {author} {\bibfnamefont {P.}~\bibnamefont
  {Guillon}},\ }\bibfield  {title} {\bibinfo {title} {Dielectric resonators},\
  }\href@noop {} {\bibfield  {journal} {\bibinfo  {journal} {Norwood, MA,
  Artech House, Inc., 1986, 547 p. No individual items are abstracted in this
  volume.}\ } (\bibinfo {year} {1986})}\BibitemShut {NoStop}%
\bibitem [{\citenamefont {Volyanskiy}\ \emph {et~al.}(2008)\citenamefont
  {Volyanskiy}, \citenamefont {Cussey}, \citenamefont {Tavernier},
  \citenamefont {Salzenstein}, \citenamefont {Sauvage}, \citenamefont
  {Larger},\ and\ \citenamefont {Rubiola}}]{volyanskiy2008applications}%
  \BibitemOpen
  \bibfield  {author} {\bibinfo {author} {\bibfnamefont {K.}~\bibnamefont
  {Volyanskiy}}, \bibinfo {author} {\bibfnamefont {J.}~\bibnamefont {Cussey}},
  \bibinfo {author} {\bibfnamefont {H.}~\bibnamefont {Tavernier}}, \bibinfo
  {author} {\bibfnamefont {P.}~\bibnamefont {Salzenstein}}, \bibinfo {author}
  {\bibfnamefont {G.}~\bibnamefont {Sauvage}}, \bibinfo {author} {\bibfnamefont
  {L.}~\bibnamefont {Larger}},\ and\ \bibinfo {author} {\bibfnamefont
  {E.}~\bibnamefont {Rubiola}},\ }\bibfield  {title} {\bibinfo {title}
  {Applications of the optical fiber to the generation and measurement of
  low-phase-noise microwave signals},\ }\href
  {https://doi.org/10.1364/JOSAB.25.002140} {\bibfield  {journal} {\bibinfo
  {journal} {JOSA B}\ }\textbf {\bibinfo {volume} {25}},\ \bibinfo {pages}
  {2140} (\bibinfo {year} {2008})}\BibitemShut {NoStop}%
\bibitem [{\citenamefont {{Lakin}}\ \emph {et~al.}(1993)\citenamefont
  {{Lakin}}, \citenamefont {{Kline}},\ and\ \citenamefont
  {{McCarron}}}]{Lakin1993}%
  \BibitemOpen
  \bibfield  {author} {\bibinfo {author} {\bibfnamefont {K.~M.}\ \bibnamefont
  {{Lakin}}}, \bibinfo {author} {\bibfnamefont {G.~R.}\ \bibnamefont
  {{Kline}}},\ and\ \bibinfo {author} {\bibfnamefont {K.~T.}\ \bibnamefont
  {{McCarron}}},\ }\bibfield  {title} {\bibinfo {title} {High-q microwave
  acoustic resonators and filters},\ }\href {https://doi.org/10.1109/22.260698}
  {\bibfield  {journal} {\bibinfo  {journal} {IEEE Transactions on Microwave
  Theory and Techniques}\ }\textbf {\bibinfo {volume} {41}},\ \bibinfo {pages}
  {2139} (\bibinfo {year} {1993})}\BibitemShut {NoStop}%
\bibitem [{\citenamefont {Bankowski}\ \emph {et~al.}(2015)\citenamefont
  {Bankowski}, \citenamefont {Meitzler}, \citenamefont {Khymyn}, \citenamefont
  {Tiberkevich}, \citenamefont {Slavin},\ and\ \citenamefont
  {Tang}}]{Bankowski2015}%
  \BibitemOpen
  \bibfield  {author} {\bibinfo {author} {\bibfnamefont {E.}~\bibnamefont
  {Bankowski}}, \bibinfo {author} {\bibfnamefont {T.}~\bibnamefont {Meitzler}},
  \bibinfo {author} {\bibfnamefont {R.~S.}\ \bibnamefont {Khymyn}}, \bibinfo
  {author} {\bibfnamefont {V.~S.}\ \bibnamefont {Tiberkevich}}, \bibinfo
  {author} {\bibfnamefont {A.~N.}\ \bibnamefont {Slavin}},\ and\ \bibinfo
  {author} {\bibfnamefont {H.~X.}\ \bibnamefont {Tang}},\ }\bibfield  {title}
  {\bibinfo {title} {Magnonic crystal as a delay line for low-noise
  auto-oscillators},\ }\href {https://doi.org/10.1063/1.4931758} {\bibfield
  {journal} {\bibinfo  {journal} {Applied Physics Letters}\ }\textbf {\bibinfo
  {volume} {107}},\ \bibinfo {pages} {122409} (\bibinfo {year}
  {2015})}\BibitemShut {NoStop}%
\bibitem [{\citenamefont {Vitko}\ \emph {et~al.}(2018)\citenamefont {Vitko},
  \citenamefont {Nikitin}, \citenamefont {Ustinov},\ and\ \citenamefont
  {Kalinikos}}]{VitkoOptoElec2018}%
  \BibitemOpen
  \bibfield  {author} {\bibinfo {author} {\bibfnamefont {V.}~\bibnamefont
  {Vitko}}, \bibinfo {author} {\bibfnamefont {A.}~\bibnamefont {Nikitin}},
  \bibinfo {author} {\bibfnamefont {A.}~\bibnamefont {Ustinov}},\ and\ \bibinfo
  {author} {\bibfnamefont {B.}~\bibnamefont {Kalinikos}},\ }\bibfield  {title}
  {\bibinfo {title} {A theoretical model of dual tunable optoelectronic
  oscillator},\ }\href {https://doi.org/10.1109/IVForum.2017.8246098}
  {\bibfield  {journal} {\bibinfo  {journal} {Journal of Physics: Conference
  Series}\ }\textbf {\bibinfo {volume} {1038}},\ \bibinfo {pages} {012106}
  (\bibinfo {year} {2018})}\BibitemShut {NoStop}%
\bibitem [{\citenamefont {Comstock}(1965)}]{Comstock1965}%
  \BibitemOpen
  \bibfield  {author} {\bibinfo {author} {\bibfnamefont {R.}~\bibnamefont
  {Comstock}},\ }\bibfield  {title} {\bibinfo {title} {Magnetoelastic coupling
  constants of the ferrites and garnets},\ }\href
  {https://doi.org/10.1109/PROC.1965.4263} {\bibfield  {journal} {\bibinfo
  {journal} {Proceedings of the IEEE}\ }\textbf {\bibinfo {volume} {53}},\
  \bibinfo {pages} {1508} (\bibinfo {year} {1965})}\BibitemShut {NoStop}%
\bibitem [{\citenamefont {Kittel}(1958)}]{Kittel1958}%
  \BibitemOpen
  \bibfield  {author} {\bibinfo {author} {\bibfnamefont {C.}~\bibnamefont
  {Kittel}},\ }\bibfield  {title} {\bibinfo {title} {Interaction of spin waves
  and ultrasonic waves in ferromagnetic crystals},\ }\href
  {https://doi.org/10.1103/PhysRev.110.836} {\bibfield  {journal} {\bibinfo
  {journal} {Physical Review}\ }\textbf {\bibinfo {volume} {110}},\ \bibinfo
  {pages} {836} (\bibinfo {year} {1958})}\BibitemShut {NoStop}%
\bibitem [{\citenamefont {Spencer}\ and\ \citenamefont
  {LeCraw}(1958)}]{Spencer1958}%
  \BibitemOpen
  \bibfield  {author} {\bibinfo {author} {\bibfnamefont {E.~G.}\ \bibnamefont
  {Spencer}}\ and\ \bibinfo {author} {\bibfnamefont {R.}~\bibnamefont
  {LeCraw}},\ }\bibfield  {title} {\bibinfo {title} {Magnetoacoustic resonance
  in yttrium iron garnet},\ }\href {https://doi.org/10.1103/PhysRevLett.1.241}
  {\bibfield  {journal} {\bibinfo  {journal} {Physical Review Letters}\
  }\textbf {\bibinfo {volume} {1}},\ \bibinfo {pages} {241} (\bibinfo {year}
  {1958})}\BibitemShut {NoStop}%
\bibitem [{\citenamefont {Eshbach}(1963)}]{Eshbach1963}%
  \BibitemOpen
  \bibfield  {author} {\bibinfo {author} {\bibfnamefont {J.~R.}\ \bibnamefont
  {Eshbach}},\ }\bibfield  {title} {\bibinfo {title} {Spin‐wave propagation
  and the magnetoelastic interaction in yttrium iron garnet},\ }\href
  {https://doi.org/10.1063/1.1729481} {\bibfield  {journal} {\bibinfo
  {journal} {Journal of Applied Physics}\ }\textbf {\bibinfo {volume} {34}},\
  \bibinfo {pages} {1298} (\bibinfo {year} {1963})}\BibitemShut {NoStop}%
\bibitem [{\citenamefont {Schl{\"o}mann}\ and\ \citenamefont
  {Joseph}(1964)}]{Schlomann1964}%
  \BibitemOpen
  \bibfield  {author} {\bibinfo {author} {\bibfnamefont {E.}~\bibnamefont
  {Schl{\"o}mann}}\ and\ \bibinfo {author} {\bibfnamefont {R.~I.}\ \bibnamefont
  {Joseph}},\ }\bibfield  {title} {\bibinfo {title} {Generation of spin waves
  in nonuniform magnetic fields. iii. magnetoelastic interaction},\ }\href
  {https://doi.org/10.1063/1.1702867} {\bibfield  {journal} {\bibinfo
  {journal} {Journal of Applied Physics}\ }\textbf {\bibinfo {volume} {35}},\
  \bibinfo {pages} {2382} (\bibinfo {year} {1964})}\BibitemShut {NoStop}%
\bibitem [{\citenamefont {Strauss}(1965)}]{Strauss1965}%
  \BibitemOpen
  \bibfield  {author} {\bibinfo {author} {\bibfnamefont {W.}~\bibnamefont
  {Strauss}},\ }\bibfield  {title} {\bibinfo {title} {Magnetoelastic waves in
  yttrium iron garnet},\ }\href {https://doi.org/10.1063/1.1713856} {\bibfield
  {journal} {\bibinfo  {journal} {Journal of Applied Physics}\ }\textbf
  {\bibinfo {volume} {36}},\ \bibinfo {pages} {118} (\bibinfo {year}
  {1965})}\BibitemShut {NoStop}%
\bibitem [{\citenamefont {Auld}\ \emph {et~al.}(1968)\citenamefont {Auld},
  \citenamefont {Collins},\ and\ \citenamefont {Webb}}]{Auld1968}%
  \BibitemOpen
  \bibfield  {author} {\bibinfo {author} {\bibfnamefont {B.}~\bibnamefont
  {Auld}}, \bibinfo {author} {\bibfnamefont {J.}~\bibnamefont {Collins}},\ and\
  \bibinfo {author} {\bibfnamefont {D.}~\bibnamefont {Webb}},\ }\bibfield
  {title} {\bibinfo {title} {Excitation of magnetoelastic waves in yig delay
  lines},\ }\href {https://doi.org/10.1063/1.1656401} {\bibfield  {journal}
  {\bibinfo  {journal} {Journal of Applied Physics}\ }\textbf {\bibinfo
  {volume} {39}},\ \bibinfo {pages} {1598} (\bibinfo {year}
  {1968})}\BibitemShut {NoStop}%
\bibitem [{\citenamefont {Rezende}\ and\ \citenamefont
  {Morgenthaler}(1969)}]{Rezende1969}%
  \BibitemOpen
  \bibfield  {author} {\bibinfo {author} {\bibfnamefont {S.~M.}\ \bibnamefont
  {Rezende}}\ and\ \bibinfo {author} {\bibfnamefont {F.~R.}\ \bibnamefont
  {Morgenthaler}},\ }\bibfield  {title} {\bibinfo {title} {Magnetoelastic waves
  in time‐varying magnetic fields. i. theory},\ }\href
  {https://doi.org/10.1063/1.1657433} {\bibfield  {journal} {\bibinfo
  {journal} {Journal of Applied Physics}\ }\textbf {\bibinfo {volume} {40}},\
  \bibinfo {pages} {524} (\bibinfo {year} {1969})}\BibitemShut {NoStop}%
\bibitem [{\citenamefont {Matthews}\ and\ \citenamefont {van~de
  Vaart}(1969)}]{Matthews1969}%
  \BibitemOpen
  \bibfield  {author} {\bibinfo {author} {\bibfnamefont {H.}~\bibnamefont
  {Matthews}}\ and\ \bibinfo {author} {\bibfnamefont {H.}~\bibnamefont {van~de
  Vaart}},\ }\bibfield  {title} {\bibinfo {title} {Magnetoelastic love waves},\
  }\href {https://doi.org/10.1063/1.1652865} {\bibfield  {journal} {\bibinfo
  {journal} {Applied Physics Letters}\ }\textbf {\bibinfo {volume} {15}},\
  \bibinfo {pages} {373} (\bibinfo {year} {1969})}\BibitemShut {NoStop}%
\bibitem [{\citenamefont {{Gulyaev}}\ \emph {et~al.}(1981)\citenamefont
  {{Gulyaev}}, \citenamefont {{ZilBerman}}, \citenamefont {{Kazakov}},
  \citenamefont {{Sysoev}}, \citenamefont {{Tikhonov}}, \citenamefont
  {{Filimonov}}, \citenamefont {{Nam}},\ and\ \citenamefont
  {{Khe}}}]{Gulyaev1981}%
  \BibitemOpen
  \bibfield  {author} {\bibinfo {author} {\bibfnamefont {Y.~V.}\ \bibnamefont
  {{Gulyaev}}}, \bibinfo {author} {\bibfnamefont {P.~E.}\ \bibnamefont
  {{ZilBerman}}}, \bibinfo {author} {\bibfnamefont {G.~T.}\ \bibnamefont
  {{Kazakov}}}, \bibinfo {author} {\bibfnamefont {V.~G.}\ \bibnamefont
  {{Sysoev}}}, \bibinfo {author} {\bibfnamefont {V.~V.}\ \bibnamefont
  {{Tikhonov}}}, \bibinfo {author} {\bibfnamefont {Y.~A.}\ \bibnamefont
  {{Filimonov}}}, \bibinfo {author} {\bibfnamefont {B.~P.}\ \bibnamefont
  {{Nam}}},\ and\ \bibinfo {author} {\bibfnamefont {A.~S.}\ \bibnamefont
  {{Khe}}},\ }\bibfield  {title} {\bibinfo {title} {Observation of fast
  magnetoelastic waves in thin yttrium-iron garnet wafers and epitaxial
  films},\ }\href@noop {} {\bibfield  {journal} {\bibinfo  {journal} {Soviet
  Journal of Experimental and Theoretical Physics Letters}\ }\textbf {\bibinfo
  {volume} {34}},\ \bibinfo {pages} {477} (\bibinfo {year} {1981})}\BibitemShut
  {NoStop}%
\bibitem [{\citenamefont {Kazakov}\ \emph {et~al.}(1983)\citenamefont
  {Kazakov}, \citenamefont {Tikhonov},\ and\ \citenamefont
  {Zilberman}}]{Kazakov1983}%
  \BibitemOpen
  \bibfield  {author} {\bibinfo {author} {\bibfnamefont {G.}~\bibnamefont
  {Kazakov}}, \bibinfo {author} {\bibfnamefont {V.}~\bibnamefont {Tikhonov}},\
  and\ \bibinfo {author} {\bibfnamefont {P.}~\bibnamefont {Zilberman}},\
  }\bibfield  {title} {\bibinfo {title} {Magneto-dipole and elastic wave
  resonance interaction in yig plates and films},\ }\href@noop {} {\bibfield
  {journal} {\bibinfo  {journal} {Fizika Tverdogo Tela}\ }\textbf {\bibinfo
  {volume} {25}},\ \bibinfo {pages} {2307} (\bibinfo {year}
  {1983})}\BibitemShut {NoStop}%
\bibitem [{\citenamefont {Zilberman}\ \emph {et~al.}(1985)\citenamefont
  {Zilberman}, \citenamefont {Kazakov},\ and\ \citenamefont
  {Tikhonov}}]{Zilberman1985}%
  \BibitemOpen
  \bibfield  {author} {\bibinfo {author} {\bibfnamefont {P.}~\bibnamefont
  {Zilberman}}, \bibinfo {author} {\bibfnamefont {G.}~\bibnamefont {Kazakov}},\
  and\ \bibinfo {author} {\bibfnamefont {V.}~\bibnamefont {Tikhonov}},\
  }\bibfield  {title} {\bibinfo {title} {Self-modulation of fast magnetoelastic
  waves in yttrium iron garnet films},\ }\href@noop {} {\bibfield  {journal}
  {\bibinfo  {journal} {Technical Physics Letters}\ }\textbf {\bibinfo {volume}
  {11}},\ \bibinfo {pages} {769} (\bibinfo {year} {1985})}\BibitemShut
  {NoStop}%
\bibitem [{\citenamefont {Gulyaev}\ and\ \citenamefont
  {ZilBerman}(1988)}]{Gulyaev1988}%
  \BibitemOpen
  \bibfield  {author} {\bibinfo {author} {\bibfnamefont {Y.~V.}\ \bibnamefont
  {Gulyaev}}\ and\ \bibinfo {author} {\bibfnamefont {P.}~\bibnamefont
  {ZilBerman}},\ }\bibfield  {title} {\bibinfo {title} {Magnetoelastic waves in
  ferromagnet plates and films},\ }\href@noop {} {\bibfield  {journal}
  {\bibinfo  {journal} {Soviet Physics Journal}\ }\textbf {\bibinfo {volume}
  {31}},\ \bibinfo {pages} {860} (\bibinfo {year} {1988})}\BibitemShut
  {NoStop}%
\bibitem [{\citenamefont {Weiler}\ \emph {et~al.}(2012)\citenamefont {Weiler},
  \citenamefont {Huebl}, \citenamefont {Goerg}, \citenamefont {Czeschka},
  \citenamefont {Gross},\ and\ \citenamefont {Goennenwein}}]{Weiler2012}%
  \BibitemOpen
  \bibfield  {author} {\bibinfo {author} {\bibfnamefont {M.}~\bibnamefont
  {Weiler}}, \bibinfo {author} {\bibfnamefont {H.}~\bibnamefont {Huebl}},
  \bibinfo {author} {\bibfnamefont {F.~S.}\ \bibnamefont {Goerg}}, \bibinfo
  {author} {\bibfnamefont {F.~D.}\ \bibnamefont {Czeschka}}, \bibinfo {author}
  {\bibfnamefont {R.}~\bibnamefont {Gross}},\ and\ \bibinfo {author}
  {\bibfnamefont {S.~T.~B.}\ \bibnamefont {Goennenwein}},\ }\bibfield  {title}
  {\bibinfo {title} {Spin pumping with coherent elastic waves},\ }\href
  {https://doi.org/10.1103/PhysRevLett.108.176601} {\bibfield  {journal}
  {\bibinfo  {journal} {Phys. Rev. Lett.}\ }\textbf {\bibinfo {volume} {108}},\
  \bibinfo {pages} {176601} (\bibinfo {year} {2012})}\BibitemShut {NoStop}%
\bibitem [{\citenamefont {{Chowdhury}}\ \emph {et~al.}(2015)\citenamefont
  {{Chowdhury}}, \citenamefont {{Dhagat}},\ and\ \citenamefont
  {{Jander}}}]{Chowdhury2015}%
  \BibitemOpen
  \bibfield  {author} {\bibinfo {author} {\bibfnamefont {P.}~\bibnamefont
  {{Chowdhury}}}, \bibinfo {author} {\bibfnamefont {P.}~\bibnamefont
  {{Dhagat}}},\ and\ \bibinfo {author} {\bibfnamefont {A.}~\bibnamefont
  {{Jander}}},\ }\bibfield  {title} {\bibinfo {title} {Parametric amplification
  of spin waves using acoustic waves},\ }\href
  {https://doi.org/10.1109/TMAG.2015.2445791} {\bibfield  {journal} {\bibinfo
  {journal} {IEEE Transactions on Magnetics}\ }\textbf {\bibinfo {volume}
  {51}},\ \bibinfo {pages} {1} (\bibinfo {year} {2015})}\BibitemShut {NoStop}%
\bibitem [{\citenamefont {Khivintsev}\ \emph {et~al.}(2018)\citenamefont
  {Khivintsev}, \citenamefont {Sakharov}, \citenamefont {Vysotskii},
  \citenamefont {Filimonov},\ and\ \citenamefont {Nikitov}}]{Khivintsev2018}%
  \BibitemOpen
  \bibfield  {author} {\bibinfo {author} {\bibfnamefont {Y.~V.}\ \bibnamefont
  {Khivintsev}}, \bibinfo {author} {\bibfnamefont {V.~K.}\ \bibnamefont
  {Sakharov}}, \bibinfo {author} {\bibfnamefont {S.~L.}\ \bibnamefont
  {Vysotskii}}, \bibinfo {author} {\bibfnamefont {A.~I.}\ \bibnamefont
  {Filimonov}, \bibfnamefont {Yu. A.and~Stognii}},\ and\ \bibinfo {author}
  {\bibfnamefont {S.~A.}\ \bibnamefont {Nikitov}},\ }\bibfield  {title}
  {\bibinfo {title} {Magnetoelastic waves in submicron yttrium--iron garnet
  films manufactured by means of ion-beam sputtering onto gadolinium--gallium
  garnet substrates},\ }\href {https://doi.org/10.1134/S1063784218070162}
  {\bibfield  {journal} {\bibinfo  {journal} {Technical Physics}\ }\textbf
  {\bibinfo {volume} {63}},\ \bibinfo {pages} {1029} (\bibinfo {year}
  {2018})}\BibitemShut {NoStop}%
\bibitem [{\citenamefont {Polzikova}\ \emph {et~al.}(2013)\citenamefont
  {Polzikova}, \citenamefont {Alekseev}, \citenamefont {Kotelyanskii},
  \citenamefont {Raevskiy},\ and\ \citenamefont {Fetisov}}]{Polzikova2013}%
  \BibitemOpen
  \bibfield  {author} {\bibinfo {author} {\bibfnamefont {N.}~\bibnamefont
  {Polzikova}}, \bibinfo {author} {\bibfnamefont {S.}~\bibnamefont {Alekseev}},
  \bibinfo {author} {\bibfnamefont {I.}~\bibnamefont {Kotelyanskii}}, \bibinfo
  {author} {\bibfnamefont {A.}~\bibnamefont {Raevskiy}},\ and\ \bibinfo
  {author} {\bibfnamefont {Y.}~\bibnamefont {Fetisov}},\ }\bibfield  {title}
  {\bibinfo {title} {Magnetic field tunable acoustic resonator with
  ferromagnetic-ferroelectric layered structure},\ }\href
  {https://doi.org/10.1063/1.4793774} {\bibfield  {journal} {\bibinfo
  {journal} {Journal of Applied Physics}\ }\textbf {\bibinfo {volume} {113}},\
  \bibinfo {pages} {17C704} (\bibinfo {year} {2013})}\BibitemShut {NoStop}%
\bibitem [{\citenamefont {Pyataikin}\ \emph {et~al.}(2017)\citenamefont
  {Pyataikin}, \citenamefont {Polzikova}, \citenamefont {Alekseev},
  \citenamefont {Kotelyanskii}, \citenamefont {Luzanov}, \citenamefont
  {Raevskiy},\ and\ \citenamefont {Galchenkov}}]{Pyataikin2017}%
  \BibitemOpen
  \bibfield  {author} {\bibinfo {author} {\bibfnamefont {I.}~\bibnamefont
  {Pyataikin}}, \bibinfo {author} {\bibfnamefont {N.}~\bibnamefont
  {Polzikova}}, \bibinfo {author} {\bibfnamefont {S.}~\bibnamefont {Alekseev}},
  \bibinfo {author} {\bibfnamefont {I.}~\bibnamefont {Kotelyanskii}}, \bibinfo
  {author} {\bibfnamefont {V.}~\bibnamefont {Luzanov}}, \bibinfo {author}
  {\bibfnamefont {A.}~\bibnamefont {Raevskiy}},\ and\ \bibinfo {author}
  {\bibfnamefont {L.}~\bibnamefont {Galchenkov}},\ }\bibfield  {title}
  {\bibinfo {title} {Spin pumping in a composite high overtone bulk acoustic
  wave resonator},\ }\href {https://doi.org/10.3103/S1062873817080251}
  {\bibfield  {journal} {\bibinfo  {journal} {Bulletin of the Russian Academy
  of Sciences: Physics}\ }\textbf {\bibinfo {volume} {81}},\ \bibinfo {pages}
  {962} (\bibinfo {year} {2017})}\BibitemShut {NoStop}%
\bibitem [{\citenamefont {Polzikova}\ \emph {et~al.}(2019)\citenamefont
  {Polzikova}, \citenamefont {Alekseev}, \citenamefont {Luzanov},\ and\
  \citenamefont {Raevskiy}}]{Polzikova2019}%
  \BibitemOpen
  \bibfield  {author} {\bibinfo {author} {\bibfnamefont {N.}~\bibnamefont
  {Polzikova}}, \bibinfo {author} {\bibfnamefont {S.}~\bibnamefont {Alekseev}},
  \bibinfo {author} {\bibfnamefont {V.}~\bibnamefont {Luzanov}},\ and\ \bibinfo
  {author} {\bibfnamefont {A.}~\bibnamefont {Raevskiy}},\ }\bibfield  {title}
  {\bibinfo {title} {Acoustic excitation and electrical detection of spin waves
  and spin currents in hypersonic bulk waves resonator with yig/pt system},\
  }\href {https://doi.org/10.1016/j.jmmm.2019.02.007} {\bibfield  {journal}
  {\bibinfo  {journal} {Journal of Magnetism and Magnetic Materials}\ }\textbf
  {\bibinfo {volume} {479}},\ \bibinfo {pages} {38} (\bibinfo {year}
  {2019})}\BibitemShut {NoStop}%
\bibitem [{\citenamefont {Alekseev}\ \emph {et~al.}(2020)\citenamefont
  {Alekseev}, \citenamefont {Dizhur}, \citenamefont {Polzikova}, \citenamefont
  {Luzanov}, \citenamefont {Raevskiy}, \citenamefont {Orlov}, \citenamefont
  {Kotov},\ and\ \citenamefont {Nikitov}}]{Alekseev2020}%
  \BibitemOpen
  \bibfield  {author} {\bibinfo {author} {\bibfnamefont {S.~G.}\ \bibnamefont
  {Alekseev}}, \bibinfo {author} {\bibfnamefont {S.~E.}\ \bibnamefont
  {Dizhur}}, \bibinfo {author} {\bibfnamefont {N.~I.}\ \bibnamefont
  {Polzikova}}, \bibinfo {author} {\bibfnamefont {V.~A.}\ \bibnamefont
  {Luzanov}}, \bibinfo {author} {\bibfnamefont {A.~O.}\ \bibnamefont
  {Raevskiy}}, \bibinfo {author} {\bibfnamefont {A.~P.}\ \bibnamefont {Orlov}},
  \bibinfo {author} {\bibfnamefont {V.~A.}\ \bibnamefont {Kotov}},\ and\
  \bibinfo {author} {\bibfnamefont {S.~A.}\ \bibnamefont {Nikitov}},\
  }\bibfield  {title} {\bibinfo {title} {Magnons parametric pumping in bulk
  acoustic waves resonator},\ }\href {https://doi.org/10.1063/5.0022267}
  {\bibfield  {journal} {\bibinfo  {journal} {Applied Physics Letters}\
  }\textbf {\bibinfo {volume} {117}},\ \bibinfo {pages} {072408} (\bibinfo
  {year} {2020})}\BibitemShut {NoStop}%
\bibitem [{\citenamefont {{Litvinenko}}\ \emph {et~al.}(2015)\citenamefont
  {{Litvinenko}}, \citenamefont {{Sadovnikov}}, \citenamefont {{Tikhonov}},\
  and\ \citenamefont {{Nikitov}}}]{Litvinenko2015}%
  \BibitemOpen
  \bibfield  {author} {\bibinfo {author} {\bibfnamefont {A.~N.}\ \bibnamefont
  {{Litvinenko}}}, \bibinfo {author} {\bibfnamefont {A.~V.}\ \bibnamefont
  {{Sadovnikov}}}, \bibinfo {author} {\bibfnamefont {V.~V.}\ \bibnamefont
  {{Tikhonov}}},\ and\ \bibinfo {author} {\bibfnamefont {S.~A.}\ \bibnamefont
  {{Nikitov}}},\ }\bibfield  {title} {\bibinfo {title} {Brillouin light
  scattering spectroscopy of magneto-acoustic resonances in a thin-film garnet
  resonator},\ }\href {https://doi.org/10.1109/LMAG.2015.2494008} {\bibfield
  {journal} {\bibinfo  {journal} {IEEE Magnetics Letters}\ }\textbf {\bibinfo
  {volume} {6}},\ \bibinfo {pages} {1} (\bibinfo {year} {2015})}\BibitemShut
  {NoStop}%
\bibitem [{\citenamefont {An}\ \emph {et~al.}(2020)\citenamefont {An},
  \citenamefont {Litvinenko}, \citenamefont {Kohno}, \citenamefont {Fuad},
  \citenamefont {Naletov}, \citenamefont {Vila}, \citenamefont {Ebels},
  \citenamefont {de~Loubens}, \citenamefont {Hurdequint}, \citenamefont
  {Beaulieu} \emph {et~al.}}]{An2020}%
  \BibitemOpen
  \bibfield  {author} {\bibinfo {author} {\bibfnamefont {K.}~\bibnamefont
  {An}}, \bibinfo {author} {\bibfnamefont {A.}~\bibnamefont {Litvinenko}},
  \bibinfo {author} {\bibfnamefont {R.}~\bibnamefont {Kohno}}, \bibinfo
  {author} {\bibfnamefont {A.}~\bibnamefont {Fuad}}, \bibinfo {author}
  {\bibfnamefont {V.}~\bibnamefont {Naletov}}, \bibinfo {author} {\bibfnamefont
  {L.}~\bibnamefont {Vila}}, \bibinfo {author} {\bibfnamefont {U.}~\bibnamefont
  {Ebels}}, \bibinfo {author} {\bibfnamefont {G.}~\bibnamefont {de~Loubens}},
  \bibinfo {author} {\bibfnamefont {H.}~\bibnamefont {Hurdequint}}, \bibinfo
  {author} {\bibfnamefont {N.}~\bibnamefont {Beaulieu}}, \emph {et~al.},\
  }\bibfield  {title} {\bibinfo {title} {Coherent long-range transfer of
  angular momentum between magnon kittel modes by phonons},\ }\href
  {https://doi.org/10.1103/PhysRevB.101.060407} {\bibfield  {journal} {\bibinfo
   {journal} {Physical Review B}\ }\textbf {\bibinfo {volume} {101}},\ \bibinfo
  {pages} {060407} (\bibinfo {year} {2020})}\BibitemShut {NoStop}%
\bibitem [{\citenamefont {Boudot}\ \emph {et~al.}(2016)\citenamefont {Boudot},
  \citenamefont {Martin}, \citenamefont {Friedt},\ and\ \citenamefont
  {Rubiola}}]{Boudot2016}%
  \BibitemOpen
  \bibfield  {author} {\bibinfo {author} {\bibfnamefont {R.}~\bibnamefont
  {Boudot}}, \bibinfo {author} {\bibfnamefont {G.}~\bibnamefont {Martin}},
  \bibinfo {author} {\bibfnamefont {J.-M.}\ \bibnamefont {Friedt}},\ and\
  \bibinfo {author} {\bibfnamefont {E.}~\bibnamefont {Rubiola}},\ }\bibfield
  {title} {\bibinfo {title} {Frequency flicker of 2.3 ghz aln-sapphire
  high-overtone bulk acoustic resonators},\ }\href
  {https://doi.org/10.1063/1.4972102} {\bibfield  {journal} {\bibinfo
  {journal} {Journal of Applied Physics}\ }\textbf {\bibinfo {volume}
  {120(22)}},\ \bibinfo {pages} {224903} (\bibinfo {year} {2016})}\BibitemShut
  {NoStop}%
\bibitem [{\citenamefont {Yu}\ \emph {et~al.}(2009)\citenamefont {Yu},
  \citenamefont {Lee}, \citenamefont {Pang}, \citenamefont {Zhang},
  \citenamefont {Brannon}, \citenamefont {Kitching},\ and\ \citenamefont
  {Kim}}]{Yu2009}%
  \BibitemOpen
  \bibfield  {author} {\bibinfo {author} {\bibfnamefont {H.}~\bibnamefont
  {Yu}}, \bibinfo {author} {\bibfnamefont {C.}~\bibnamefont {Lee}}, \bibinfo
  {author} {\bibfnamefont {W.}~\bibnamefont {Pang}}, \bibinfo {author}
  {\bibfnamefont {H.}~\bibnamefont {Zhang}}, \bibinfo {author} {\bibfnamefont
  {A.}~\bibnamefont {Brannon}}, \bibinfo {author} {\bibfnamefont
  {J.}~\bibnamefont {Kitching}},\ and\ \bibinfo {author} {\bibfnamefont
  {E.~S.}\ \bibnamefont {Kim}},\ }\bibfield  {title} {\bibinfo {title}
  {Hbar-based 3.6 ghz oscillator with low power consumption and low phase
  noise},\ }\href {https://doi.org/10.1109/TUFFC.2009.1050} {\bibfield
  {journal} {\bibinfo  {journal} {IEEE Trans. Ultrason. Ferroelectr. Freq.
  Control}\ }\textbf {\bibinfo {volume} {56(2)}},\ \bibinfo {pages} {400}
  (\bibinfo {year} {2009})}\BibitemShut {NoStop}%
\bibitem [{\citenamefont {Verba}\ \emph {et~al.}(2018)\citenamefont {Verba},
  \citenamefont {Lisenkov}, \citenamefont {Krivorotov}, \citenamefont
  {Tiberkevich},\ and\ \citenamefont {Slavin}}]{Verba2018}%
  \BibitemOpen
  \bibfield  {author} {\bibinfo {author} {\bibfnamefont {R.}~\bibnamefont
  {Verba}}, \bibinfo {author} {\bibfnamefont {I.}~\bibnamefont {Lisenkov}},
  \bibinfo {author} {\bibfnamefont {I.}~\bibnamefont {Krivorotov}}, \bibinfo
  {author} {\bibfnamefont {V.}~\bibnamefont {Tiberkevich}},\ and\ \bibinfo
  {author} {\bibfnamefont {A.}~\bibnamefont {Slavin}},\ }\bibfield  {title}
  {\bibinfo {title} {Nonreciprocal surface acoustic waves in multilayers with
  magnetoelastic and interfacial dzyaloshinskii-moriya interactions},\ }\href
  {https://doi.org/10.1103/PhysRevApplied.9.064014} {\bibfield  {journal}
  {\bibinfo  {journal} {Physical Review Applied}\ }\textbf {\bibinfo {volume}
  {9}},\ \bibinfo {pages} {064014} (\bibinfo {year} {2018})}\BibitemShut
  {NoStop}%
\bibitem [{\citenamefont {Lisenkov}\ \emph {et~al.}(2019)\citenamefont
  {Lisenkov}, \citenamefont {Jander},\ and\ \citenamefont
  {Dhagat}}]{Lisenkov2019}%
  \BibitemOpen
  \bibfield  {author} {\bibinfo {author} {\bibfnamefont {I.}~\bibnamefont
  {Lisenkov}}, \bibinfo {author} {\bibfnamefont {A.}~\bibnamefont {Jander}},\
  and\ \bibinfo {author} {\bibfnamefont {P.}~\bibnamefont {Dhagat}},\
  }\bibfield  {title} {\bibinfo {title} {Magnetoelastic parametric
  instabilities of localized spin waves induced by traveling elastic waves},\
  }\href {https://doi.org/10.1103/PhysRevB.99.184433} {\bibfield  {journal}
  {\bibinfo  {journal} {Physical Review B}\ }\textbf {\bibinfo {volume} {99}},\
  \bibinfo {pages} {184433} (\bibinfo {year} {2019})}\BibitemShut {NoStop}%
\bibitem [{\citenamefont {Chikazumi}(2009)}]{chikazumi2009physics}%
  \BibitemOpen
  \bibfield  {author} {\bibinfo {author} {\bibfnamefont {S.}~\bibnamefont
  {Chikazumi}},\ }\href {https://books.google.fr/books?id=AZVfuxXF2GsC} {\emph
  {\bibinfo {title} {Physics of Ferromagnetism}}},\ International Series of
  Monographs on Physics\ (\bibinfo  {publisher} {OUP Oxford},\ \bibinfo {year}
  {2009})\BibitemShut {NoStop}%
\bibitem [{\citenamefont {Brataas}\ \emph {et~al.}(2020)\citenamefont
  {Brataas}, \citenamefont {van Wees}, \citenamefont {Klein}, \citenamefont
  {de~Loubens},\ and\ \citenamefont {Viret}}]{brataas2020spin}%
  \BibitemOpen
  \bibfield  {author} {\bibinfo {author} {\bibfnamefont {A.}~\bibnamefont
  {Brataas}}, \bibinfo {author} {\bibfnamefont {B.}~\bibnamefont {van Wees}},
  \bibinfo {author} {\bibfnamefont {O.}~\bibnamefont {Klein}}, \bibinfo
  {author} {\bibfnamefont {G.}~\bibnamefont {de~Loubens}},\ and\ \bibinfo
  {author} {\bibfnamefont {M.}~\bibnamefont {Viret}},\ }\bibfield  {title}
  {\bibinfo {title} {Spin insulatronics},\ }\href
  {https://doi.org/10.1016/j.physrep.2020.08.006} {\bibfield  {journal}
  {\bibinfo  {journal} {Physics Reports}\ }\textbf {\bibinfo {volume} {885}},\
  \bibinfo {pages} {1} (\bibinfo {year} {2020})}\BibitemShut {NoStop}%
\bibitem [{\citenamefont {Borovik-Romanov}\ and\ \citenamefont
  {Sinha}(2012)}]{borovik2012spin}%
  \BibitemOpen
  \bibfield  {author} {\bibinfo {author} {\bibfnamefont {A.}~\bibnamefont
  {Borovik-Romanov}}\ and\ \bibinfo {author} {\bibfnamefont {S.}~\bibnamefont
  {Sinha}},\ }\href {https://books.google.fr/books?id=ypijB-Ku4moC} {\emph
  {\bibinfo {title} {Spin Waves and Magnetic Excitations}}},\ ISSN\ (\bibinfo
  {publisher} {Elsevier Science},\ \bibinfo {year} {2012})\BibitemShut
  {NoStop}%
\bibitem [{\citenamefont {Kajfez}\ and\ \citenamefont
  {Hwan}(1984)}]{Kajfez1984}%
  \BibitemOpen
  \bibfield  {author} {\bibinfo {author} {\bibfnamefont {D.}~\bibnamefont
  {Kajfez}}\ and\ \bibinfo {author} {\bibfnamefont {E.~J.}\ \bibnamefont
  {Hwan}},\ }\bibfield  {title} {\bibinfo {title} {Q-factor measurement with
  network analyzer},\ }\href {https://doi.org/10.1109/TMTT.1984.1132751}
  {\bibfield  {journal} {\bibinfo  {journal} {IEEE transactions on microwave
  theory and techniques}\ }\textbf {\bibinfo {volume} {32}},\ \bibinfo {pages}
  {666} (\bibinfo {year} {1984})}\BibitemShut {NoStop}%
\bibitem [{\citenamefont {Litvinenko}\ \emph {et~al.}(2018)\citenamefont
  {Litvinenko}, \citenamefont {Grishin}, \citenamefont {Sharaevskii},
  \citenamefont {Tikhonov},\ and\ \citenamefont
  {Nikitov}}]{litvinenko2018chaotic}%
  \BibitemOpen
  \bibfield  {author} {\bibinfo {author} {\bibfnamefont {A.}~\bibnamefont
  {Litvinenko}}, \bibinfo {author} {\bibfnamefont {S.}~\bibnamefont {Grishin}},
  \bibinfo {author} {\bibfnamefont {Y.~P.}\ \bibnamefont {Sharaevskii}},
  \bibinfo {author} {\bibfnamefont {V.}~\bibnamefont {Tikhonov}},\ and\
  \bibinfo {author} {\bibfnamefont {S.}~\bibnamefont {Nikitov}},\ }\bibfield
  {title} {\bibinfo {title} {A chaotic magnetoacoustic oscillator with delay
  and bistability},\ }\href {https://doi.org/10.1134/S1063785018030215}
  {\bibfield  {journal} {\bibinfo  {journal} {Technical Physics Letters}\
  }\textbf {\bibinfo {volume} {44}},\ \bibinfo {pages} {263} (\bibinfo {year}
  {2018})}\BibitemShut {NoStop}%
\end{thebibliography}%
%\bibitem{sample1} This is a sample bibitem

\end{document}